\documentclass[10pt,a4paper,reprint,aps]{revtex4-2}
\usepackage[utf8]{inputenc}
\usepackage{amsmath}
\usepackage{amsfonts}
\usepackage[margin=0.5in]{geometry}
\usepackage{amssymb}
\usepackage{graphicx}
\usepackage{xcolor}
\usepackage{caption}
\usepackage{subcaption}
\usepackage{color, colortbl}
\definecolor{Gray}{gray}{0.8}
\newcommand{\avg}[1]{\langle #1 \rangle}

\begin{document}

\title{\textbf{Nature of barriers determine first passage times in heterogeneous media}}

\author{Moumita Dasgupta,\textit{$^{1,\ast,\dag}$} Sougata Guha,\textit{$^{2,3,\dag}$} Leon Armbruster,\textit{$^{1}$}, Dibyendu Das\textit{$^{2}$} and Mithun K. Mitra\textit{$^{2, {\ast}}$} \\
\small{\emph{$^{1}$Department of Physics, Augsburg University, MN 55454, USA; Email: dasgupta@augsburg.edu\\
$^{2}$Department of Physics, IIT Bombay, Mumbai 400076, India; Email: mithun@phy.iitb.ac.in \\
$^{3}$INFN Napoli, Complesso Universitario di Monte S. Angelo, 80126 Napoli, Italy\\
$^\ast$Corresponding authors : dasgupta@augsburg.edu; mithun@phy.iitb.ac.in\\
$^\dag$These authors contributed equally}}}

\begin{abstract}

Intuition suggests that passage times across a region increases with the number of barriers along the path. Can this fail depending on the nature of the barrier?  To probe this fundamental question, we exactly solve for the first passage time in general $d$-dimensions for diffusive transport through a spatially patterned array of obstacles – either \emph{entropic} or \emph{energetic}, depending on the nature of the obstacles. For energetic barriers, we show that first passage times vary non-monotonically with the number of barriers, while for entropic barriers it increases monotonically. This non-monotonicity for energetic barriers further reflects in the behaviour of effective diffusivity as well. We then design a simple experiment where a robotic bug navigates a heterogeneous environment through a spatially patterned array of obstacles to validate our predictions.  Finally, using numerical simulations, we show that this non-monotonic behaviour for energetic barriers is general and extends to even super-diffusive transport.

\end{abstract}

\maketitle




\section{Introduction}
Random walks arise in multiple physical, geological, biological and ecological contexts as simple models of stochastic transport \cite{gardiner1985handbook,redner2001book,metzler2004jphysa,benichou2011revmodphys}. In particular, random walks have been used to model stochastic transport inside cells. Some well-known examples are motor proteins carrying cargo on networks of microtubules \cite{klumpp2005PNAS,muller2008PNAS,bressloff2013revmodphys}, RNA polymerase moving on DNA during transcription and backtracking \cite{roldan2016pre,guthold1999biophysj,sahoo2013JPhysCondMatt}, loop extrusion of DNA through motion of cohesin and condensin proteins \cite{maji2020biophysj,davidson2019science,banigan2020curropcellbio,ganji2018science,banigan2020elife,golfier2020elife,goloborodko2016elife}, and motion of transcription factors on DNA as it searches for binding sites \cite{kolomeisky2011pccp,van-zon2006biophysj}. \\

For physical and biological systems, often the environment in which diffusion happens is complex, characterised by heterogeneous media with frequent obstacles \cite{souza2015, Watt2006, bhattacharjee2019natcomm, dix2008annrevbiophys,goychuk2014plosone}. Inside the cell, cytoplasmic crowding can hinder transport processes. In addition, barriers may appear in the form of steric hindrance, for example, by nucleosomes to the motion of RNA polymerase during transcription \cite{jin2010natstrucmolbio,kireeva2005molcell} and to cohesin proteins during chromatin looping \cite{Stigler2016cellrep}, or by membrane furrows to diffusing morphogens in early Drosophila embryos \cite{daniels2012PNAS,rikhy2015bioopen}. Obstacles can thus affect timescales of critical biological processes -- namely transcription completion and DNA repair times, chromatin looping times, morphogen gradient formation times in embryogenesis, etc. 


In the context of such random motion through heterogeneous media, a relevant question is how effective transport coefficients arise through the interaction of the walker with the environment \cite{benichou2014physrep}. Another interesting question is completion times of transport when it reaches a target for the first time, referred to in the literature as first passage times \cite{redner2001book,bressloff2013revmodphys,chou2014first,iyerbiswas2016first,bray2013AdvPhys}. Estimating these first passage times are crucial to understanding timescales of biological processes \cite{kalinina2013natcellbio,nayak2020prr,Parmar2016nar,Stigler2016cellrep, AbhyudaySingh2017PNAS,Rijal2020pre, Rijal2022prl, Iyer-Biswas2014PNAS}. 

The role of disorder and barrier in transport may sometimes be counterintuitive. Intuitively one would expect that obstacles slow down transport, suppress effective diffusivity, enhance path lengths, and increase mean and fluctuation of first passage times. Contrary to this, experimental and theoretical studies of particle transport in presence of barriers have shown unexpected non-monotonic behavior \cite{Volpe2019natcomm,Frangipane2019natcomm,chupeau2020PNAS,Biswas2020softmatter,wagner1999pnas,palyulin2012jstatmech,palyulin2013jphysa}. For example, a study of bacterial transport through a system of microscopic scatterers showed non-monotonic behavior of effective propagation distance and effective propagation speed as a function of the obstacle density \cite{Volpe2019natcomm}. Instead of obstacles increasing times spent within a confinement, they may shorten it by decreasing the total accessible surface area \cite{Frangipane2019natcomm}. Analytical calculations have suggested that introduction of finite potential barrier can accelerate protein folding \cite{wagner1999pnas} as well as first passage times of both diffusive and subdiffusive particles \cite{palyulin2012jstatmech,palyulin2013jphysa}. Further, theoretical studies of mean first passage times (MFPT) of a diffusing particle in heterogeneous media that is subdivided into multiple regions with different diffusion coefficients has shown that the exact MFPT strongly depends on the choice of position-dependent diffusivity at the interface of the two different media \cite{vaccario2015prl}. For a two-domain system with diffusivities $D_1$ and $D_2$, the MFPT can also show an interesting non-monotonic behaviour as a function of $D_1/D_2$ depending on the starting position of the RW \cite{godec2015pre}.

In this paper, we explore the MFPT in heterogeneous environments having multiple barrier regions in general $d$-dimensional geometries. We obtain exact analytic expressions for the MFPT and show that, that depending on the \emph{nature} of the barriers, passage times may either monotonically increase as is commonly expected, or counterintuitively, are \emph{non-monotonic}, with increasing barrier numbers. Our theoretical predictions are then verified through experiments in a quasi-1D geometry.

One may broadly define two classes of barriers -- entropic vs energetic, with distinct transport properties \cite{hanggi1990revmodphys,reguera2006prl}. While energetic barriers involve activation energy in passing from one potential minima to another, entropic barriers arise due to configurational entropy differences associated with structural heterogeneity in a system \cite{Muthukumar1987jchemphys,Muthukumar1989macromolecules,bhattacharjee2019natcomm}.  In models of glass transition, the nature of barriers, whether energetic or entropic, crucially determine the nature of timescale variation \cite{Majumdar2004pre}. In our experiments, we implement the two distinct types of barriers through two different design principles. We show that while the non monotonicity in timescales of passage is observed for energetic barriers, it is not observed in the case of entropic barriers. Our exact theoretical results establish that the phenomena is not a mere artefact of the experiments, but a generic result for transport in heterogeneous media.


\section{Theory and Simulation}
\subsection*{Entropic Barriers}

\begin{figure}[t!]
    \centering
    \includegraphics[width=\linewidth]{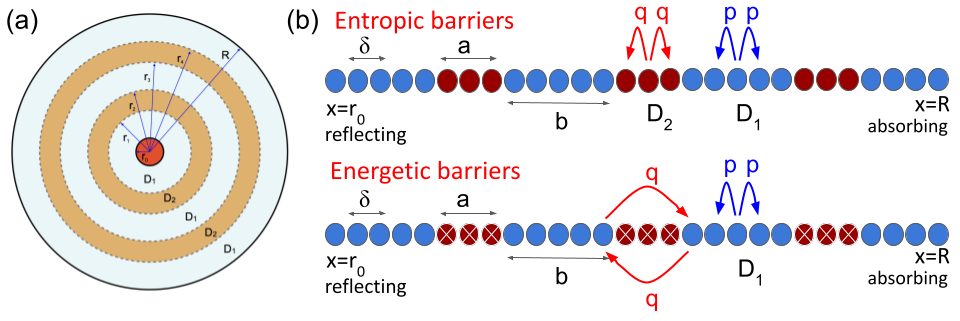}
    \caption{(a) A schematic of the heterogeneous environment with alternating diffusion coefficients $D_1$ and $D_2$. (b) The motion of a random walker in an example 1D geometry for two types of obstacles. }
    \label{fig:Fig1}
\end{figure}

We consider a $d$-dimensional hypersphere which is subdivided by $n$ barriers into $2n+1$ concentric alternating free and barrier regions (Fig. \ref{fig:Fig1}a). The simplest way to introduce barriers is through a lower diffusivity in the barrier regions, $D_2$, as compared to the free regions, $D_1$ ($>D_2$). We call this type of barrier as {\it entropic}, as the delay in navigation through the regions with lower diffusivities arises from the longer time taken to explore these regions. The $i^{th}$ barrier region is bounded by two interfaces located at $r_{2i-1}$ and $r_{2i}$. We impose a reflecting boundary near the center at $r=r_0$ and an absorbing boundary at $r=r_{2n+1}=R$. \\

The mean first passage time ($\avg{T_k}$) of a diffusing particle starting from a position in the $k^{th}$ region depends on the identity of the region, and follows
\begin{eqnarray}
    D_1 \nabla_d^2 \avg{T_{k}} &=& -1, ~ \mathrm{with}~ k=2i-1, ~ \forall i \in \{1,n+1\} \label{eq:laplace1}\\
    D_2 \nabla_d^2 \avg{T_{k}} &=& -1, ~~ \mathrm{with}~ k=2i, ~ \forall i \in \{1,n\}  \label{eq:laplace2}
\end{eqnarray}

Considering the angular symmetry of the system, we can rewrite Eqs. \ref{eq:laplace1} and \ref{eq:laplace2} as,
\begin{eqnarray}
    \frac{D_1}{r^{d-1}}\frac{\partial}{\partial r} \left(r^{d-1}\frac{\partial \avg{T_{2i-1}}}{\partial r}\right) = -1 \nonumber\\
    \frac{D_2}{r^{d-1}}\frac{\partial}{\partial r} \left(r^{d-1}\frac{\partial \avg{T_{2i}}}{\partial r}\right) = -1\nonumber
\end{eqnarray}
which has a generic solution given by,
\begin{eqnarray}
    \avg{T_{2i-1}} = \begin{cases}
    -\frac{1}{D_1}\left[\frac{r^2}{2d}+\frac{A_i r^{2-d}}{2-d}+B_i\right], & \text{for $d\neq 2$}\\
    -\frac{1}{D_1}\left[\frac{r^2}{2d}+A_i \ln r+B_i\right], & \text{for $d=2$}
    \end{cases} \label{eq:T_2i-1}
\end{eqnarray}
and \begin{eqnarray}
     \avg{T_{2i}} = \begin{cases}
    -\frac{1}{D_2}\left[\frac{r^2}{2d}+\frac{C_i r^{2-d}}{2-d}+E_i\right], & \text{for $d\neq 2$}\\
    -\frac{1}{D_2}\left[\frac{r^2}{2d}+C_i \ln r+E_i\right], & \text{for $d=2$}
    \end{cases} \label{eq:T_2i}
\end{eqnarray}
The boundary conditions at $r=r_0$ and $r=r_{2n+1}$ are given by,
\begin{eqnarray}
    \left.\frac{\partial \avg{T_1}}{\partial r}\right\vert_{r_0} = 0 \quad \text{and} \quad \left.\avg{T_{2n+1}}\right\vert_{r_{2n+1}}=0 \label{eq:boundary_condition}
\end{eqnarray}

Now, demanding that at the barrier region interfaces, the mean times as well as the corresponding “fluxes” must be continuous in any $d$-dimension, we can write the conditions,
\begin{eqnarray}
    \left.\avg{T_{2i-1}}\right\vert_{r_{2i-1}} &=& \left.\avg{T_{2i}}\right\vert_{r_{2i-1}} \label{eq:left_T}\\ D_1\left.\frac{\partial \avg{T_{2i-1}}}{\partial r}\right\vert_{r_{2i-1}} &=& D_2\left.\frac{\partial \avg{T_{2i}}}{\partial r}\right\vert_{r_{2i-1}} \label{eq:left_dT}\\
    \left.\avg{T_{2i}}\right\vert_{r_{2i}} &=& \left.\avg{T_{2i+1}}\right\vert_{r_{2i}} \label{eq:right_T}\\ D_2\left.\frac{\partial \avg{T_{2i}}}{\partial r}\right\vert_{r_{2i}} &=& D_1\left.\frac{\partial \avg{T_{2i+1}}}{\partial r}\right\vert_{r_{2i}} \label{eq:right_dT}
\end{eqnarray}

Now from Eqs. \ref{eq:T_2i-1} and \ref{eq:boundary_condition}, we get,
\begin{eqnarray}
    A_1 &=& -\frac{r_0^d}{d} \label{eq:A1}\\
    \text{and} \quad B_{n+1} &=& \begin{cases}
        -\frac{R^2}{2d} - \frac{A_{n+1}R^{2-d}}{2-d}, & \text{for $d\neq2$}\\
        -\frac{R^2}{2d} - A_{n+1}~ \ln R, & \text{for $d=2$}
    \end{cases} \label{eq:Bn1}
\end{eqnarray}

From Eqs. \ref{eq:left_dT} and \ref{eq:right_dT} one can easily obtain,
\begin{equation}
    A_i=C_i=A_{i+1}=-\frac{r_0^d}{d}, \quad \forall i \in \{1,n\}
\end{equation}

\begin{figure}[t]
    \centering
    \includegraphics[width=\linewidth]{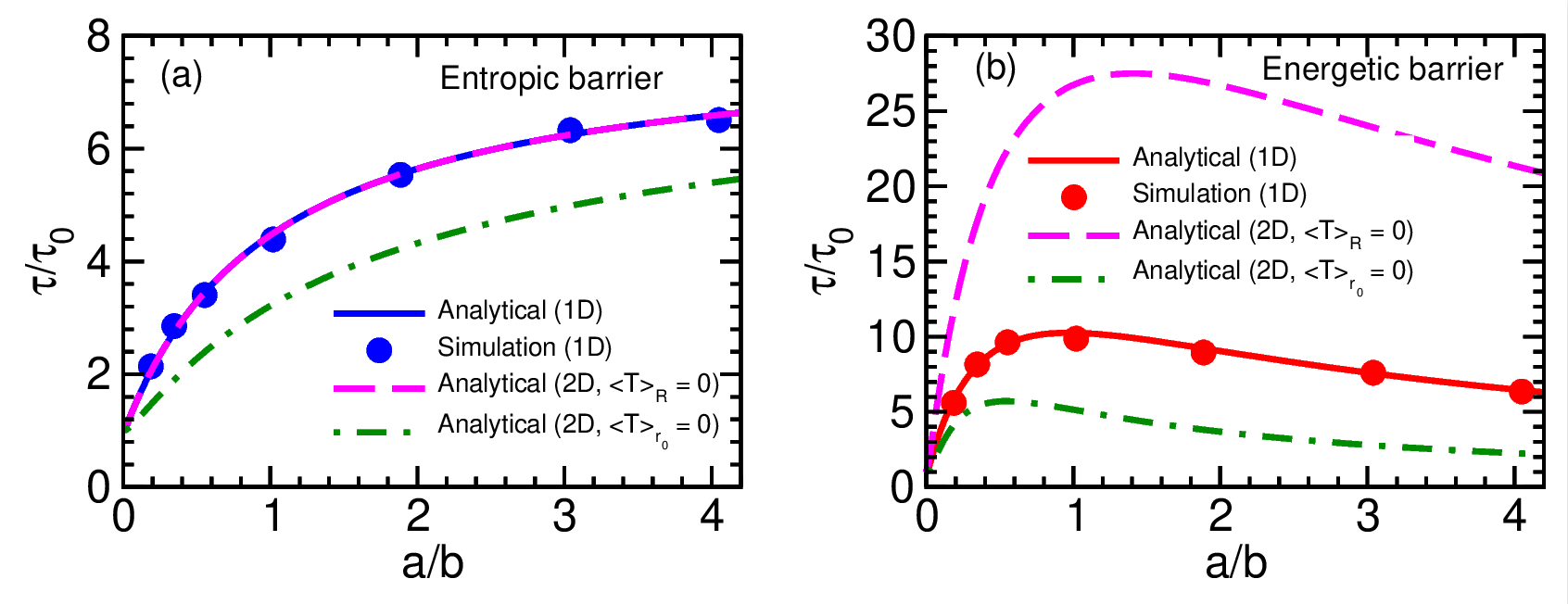}
    \caption{Scaled MFPT vs $a/b$ in case of (a) entropic barriers and (b) energetic barriers for a diffusive particle. The data is shown for the ratio $D_1/D_2 = 8$ for entropic barriers, and for $D_1/(dv_q) = 800$ for energetic barriers with $R=2001,~r_0=1$ and $a=20$. The lines are our exact expressions (Eq.~\ref{eq:scaled_MFPT} for entropic barriers, and Eq.~\ref{eq:scaled_mfpt_imp} for energetic barriers) while the points are the 1D simulation results. All simulation results in this paper are averaged over $10^4$ independent trajectories.}
    \label{fig:Fig2}
\end{figure} 

Similarly, from Eqs. \ref{eq:left_T} and \ref{eq:right_T} one gets,
\begin{eqnarray}
    B_i &=& B_{i+1} - \frac{(s-1)}{2d}\left[r_{2i}^2-r_{2i-1}^2\right] \nonumber\\ &-& \frac{(s-1)}{(2-d)}A_1 \left[r_{2i}^{2-d}-r_{2i-1}^{2-d}\right], ~ \text{for $d\neq2$} \\
    B_i = &B_{i+1}& - \frac{(s-1)}{2d}\left[r_{2i}^2-r_{2i-1}^2\right] \nonumber\\ &-&(s-1)A_1 \ln \left(\frac{r_{2i}}{r_{2i-1}}\right), \quad \text{for $d=2$} 
\end{eqnarray}
where $s=D_1/D_2 > 1$ denotes the ratio between diffusion constants in two regions.

Therefore, if the particle starts from the reflecting boundary at $r=r_0$ then the MFPT to reach $r=R$ is given by,
\begin{eqnarray}
    \tau = \avg{T_1}\vert_{r_0} = \begin{cases}
        -\frac{1}{D_1}\left[\frac{r_0^2}{2d}+ \frac{A_1r_0^{2-d}}{2-d}+B_1\right], ~\text{for $d\neq2$} \\
         -\frac{1}{D_1}\left[\frac{r_0^2}{2d}+ A_1lnr_0+B_1\right], ~ \text{for $d=2$} 
    \end{cases}
    \label{eq:MFPT_pen_d}
\end{eqnarray}
where 
\begin{eqnarray}
 B_1 = \begin{cases}
     B_{n+1} - \sum_{i=1}^{n}\frac{(s-1)}{2d}\left[r_{2i}^2-r_{2i-1}^2\right] \nonumber\\ - \sum_{i=1}^{n}\frac{(s-1)}{(2-d)}A_1 \left[r_{2i}^{2-d}-r_{2i-1}^{2-d}\right] \quad \text{for $d\neq2$} \nonumber\\
    B_{n+1} - \sum_{i=1}^{n}\frac{(s-1)}{2d}\left[r_{2i}^2-r_{2i-1}^2\right] \nonumber\\ - \sum_{i=1}^{n}(s-1)A_1 ~ln\left(\frac{r_{2i}}{r_{2i-1}}\right) \quad\quad \text{for $d=2$} \nonumber
\end{cases}
\end{eqnarray}

In Fig.~\ref{fig:Fig2}a, we plot this MFPT  as a function of the ratio of the barrier size to the size of the empty regions $(a/b)$ for $d=1,2$. Since the time in the absence of any barrier depends on the dimensionality, we scale the MFPT by the mean passage time in free space in the absence of any barriers ($\tau_0$). As the number of barriers, $n$, increases, the width of the free region decreases, and hence $a/b$ increases. We find that the MFPT increases monotonically with the ratio $a/b$, and saturates for large $a/b$ to a limiting value for all $d$. Note that the scaled MFPT is independent of dimension when the absorbing boundary is at $r=R$ (also see Fig. S4a). In $d>1$ dimension, there is an asymmetry depending on if the absorbing boundary condition is imposed on the inner ($r=r_0$) or outer ($r=R$) surface. We show this for $d=2$ by  solving the converse case where the boundary at $r=r_0$ is absorbing, and the boundary at $r=R$ is reflecting (see Supplementary Information (SI) Sec. 2.1). In this case, the scaled MFPT reduces from the case where $r=R$ is absorbing, a general trend in all dimensions $d>1$ (Fig. \ref{fig:Fig2}a and Fig. S4b). For the $1D$ case, we also perform simulations on a discrete 1D lattice with hopping probabilities $p$ and $q$ in the two regions (Fig.~\ref{fig:Fig1}b top panel), which matches exactly with our theoretical results. Thus, for these {\em entropic barriers}, consistent with our intuition, increasing barrier numbers leads to larger passage times.

In $d=1$ dimensions, Eq.~\ref{eq:MFPT_pen_d} simplifies to yield a scaled MFPT of the particle in the presence of $n$ barriers as (see SI Sec. 2.2.1 for details),
\begin{equation}
\frac{\tau}{\tau_0} = \frac{\avg{T_1}_{x=0}}{\tau_0} = 1+\frac{(s-1)(\frac{a}{b}-\frac{a}{L})}{1+\frac{a}{b}} \label{eq:scaled_MFPT}
\end{equation}
where $L=R-r_0$ is the length of the system and $s=D_1/D_2 > 1$ denotes the ratio between diffusion constants in two regions. It is easy to see from this expression that the scaled MFPT, $\tau/\tau_0$ is an increasing function of $a/b$, and hence of the number of barriers, and saturates for large $a/b$ to the limiting value $\tau = s \tau_0$. 

\subsection*{Energetic Barriers}

Are spatially varying diffusion coefficients the only way to model barrier regions? A different realisation might be to have occluded regions which are inaccessible to the RW. We consider alternating regions of blocked domains of width $a$ separated by empty regions of width $b$ (bottom panel of Fig.~\ref{fig:Fig1}b). The diffusion constant in continuum is $D_1$ for the empty regions. For the analogous discrete model (with lattice spacing $\delta$), on which simulations are performed, the unbiased forward and backward hopping rates are $p$. For hopping across a barrier from one empty region to another, the rate is $q ~(\ll p)$. The usual discrete to continuum correspondence implies $D_1 = p \delta^2$ and a barrier hopping velocity $v_q = q \delta$, in the limit $\delta \rightarrow 0$ and $p,q \rightarrow \infty$. We call these barriers {\em energetic} in an Arrhenius sense, since multiple barrier-crossing attempts are required on average to cross from one empty region to the next. We proceed to solve the problem in continuous space. For $n$ barriers, and $n+1$ free regions, the MFPT in the $k^{th}$ region follows \cite{gardiner1985handbook}


\begin{equation}
    D_1\nabla_d^2\avg{T_{k}} = -1, ~~\mathrm{with}~ k=2i-1.~ \forall i \in \{1,n+1\}
\end{equation}
In this case, the generic solution is same as Eq. \ref{eq:T_2i-1} while the boundary conditions given in Eq. \ref{eq:boundary_condition} and hence the values of $A_1$ and $B_{n+1}$ in Eqs. \ref{eq:A1} and \ref{eq:Bn1} are still valid. The matching conditions at the barrier interfaces in this case follows,
\begin{eqnarray}
    \avg{T_{2i+1}}\vert_{r_{2i}}-\avg{T_{2i-1}}\vert_{r_{2i-1}}&=&\frac{D_1}{v_q}\left.\frac{\partial \avg{T_{2i-1}}}{\partial r}\right\vert_{r_{2i-1}} \label{eq:T_2i-1_energetic}\\
    \left.\frac{\partial \avg{T_{2i-1}}}{\partial r}\right\vert_{r_{2i-1}}&=&\left.\frac{\partial \avg{T_{2i+1}}}{\partial r}\right\vert_{r_{2i}} \label{eq:dT_energetic}
\end{eqnarray}

Now, from Eq. \ref{eq:dT_energetic}, one gets,
\begin{eqnarray}
    A_{i+1}r_{2i}^{1-d}&=&A_ir_{2i-1}^{1-d}-\frac{a_i}{d}, \quad\quad \forall i \in \{1,n\}
    \label{eq:A_i+1_energetic}
\end{eqnarray}
where $a_i=r_{2i}-r_{2i-1}$ denotes the width of the $i^{th}$ barrier.

And from Eqs. \ref{eq:T_2i-1_energetic} and \ref{eq:A_i+1_energetic} one can get,
\begin{eqnarray}
    B_i &=& B_{i+1}+A_{i}r_{2i-1}^{1-d}\left[\frac{a_i}{(2-d)}-\frac{D_1}{v_q}\right] \\ &+&\frac{1}{2d}\left[r_{2i}^2-r_{2i-1}^2\right]-\frac{a_ir_{2i}}{d(2-d)}-\frac{D_1r_{2i-1}}{dv_q}, \quad \text{for $d\neq 2$} \nonumber\\
    B_i &=& B_{i+1}+A_ir_{2i-1}^{1-d}\left[\frac{\ln r_{2i}}{r_{2i}^{1-d}}-\frac{\ln r_{2i-1}}{r_{2i-1}^{1-d}}-\frac{D_1}{v_q}\right]-\frac{a_i\ln r_{2i}}{dr_{2i}^{1-d}} \nonumber\\ &+&\frac{1}{2d}\left[r_{2i}^2-r_{2i-1}^2\right]-\frac{D_1r_{2i-1}}{dv_q}, ~\quad\text{for $d=2$}
\end{eqnarray}

Therefore, if the particle starts from the reflecting boundary at $r=r_0$ then the MFPT to reach $r=R$ is given by,
\begin{eqnarray}
    \tau = \avg{T_1}\vert_{r_0} = \begin{cases}
        -\frac{1}{D_1}\left[\frac{r_0^2}{2d}+ \frac{A_1r_0^{2-d}}{2-d}+B_1\right], ~\text{for $d\neq2$} \\
         -\frac{1}{D_1}\left[\frac{r_0^2}{2d}+ A_1\ln r_0+B_1\right], ~ \text{for $d=2$} 
    \end{cases}
    \label{eq:MFPT_impen_d}
\end{eqnarray}
where 
\begin{eqnarray}
 B_1 = \begin{cases}
     B_{n+1} + \sum_{i=1}^{n}\frac{1}{2d}\left[r_{2i}^2-r_{2i-1}^2\right] \nonumber\\ +\sum_{i=1}^{n}A_ir_{2i-1}^{1-d}\left[\frac{a_i}{(2-d)}-\frac{D_1}{v_q}\right] \nonumber\\-\sum_{i=1}^{n}\frac{a_ir_{2i}}{d(2-d)}-\sum_{i=1}^{n}\frac{D_1r_{2i-1}}{dv_q}, \quad \text{for $d\neq 2$} \\
    B_{n+1} +\sum_{i=1}^{n}\frac{1}{2d}\left[r_{2i}^2-r_{2i-1}^2\right]\nonumber\\ +\sum_{i=1}^{n}A_ir_{2i-1}^{1-d}\left[\frac{\ln r_{2i}}{r_{2i}^{1-d}}-\frac{\ln r_{2i-1}}{r_{2i-1}^{1-d}}-\frac{D_1}{v_q}\right] \nonumber\\ -\sum_{i=1}^{n}\frac{a_i\ln r_{2i}}{dr_{2i}^{1-d}} -\sum_{i=1}^{n}\frac{D_1r_{2i-1}}{dv_q}, ~ \text{for $d=2$} \nonumber
\end{cases}
\end{eqnarray}

\begin{figure}[t!]
    \centering
    \includegraphics[width=\linewidth]{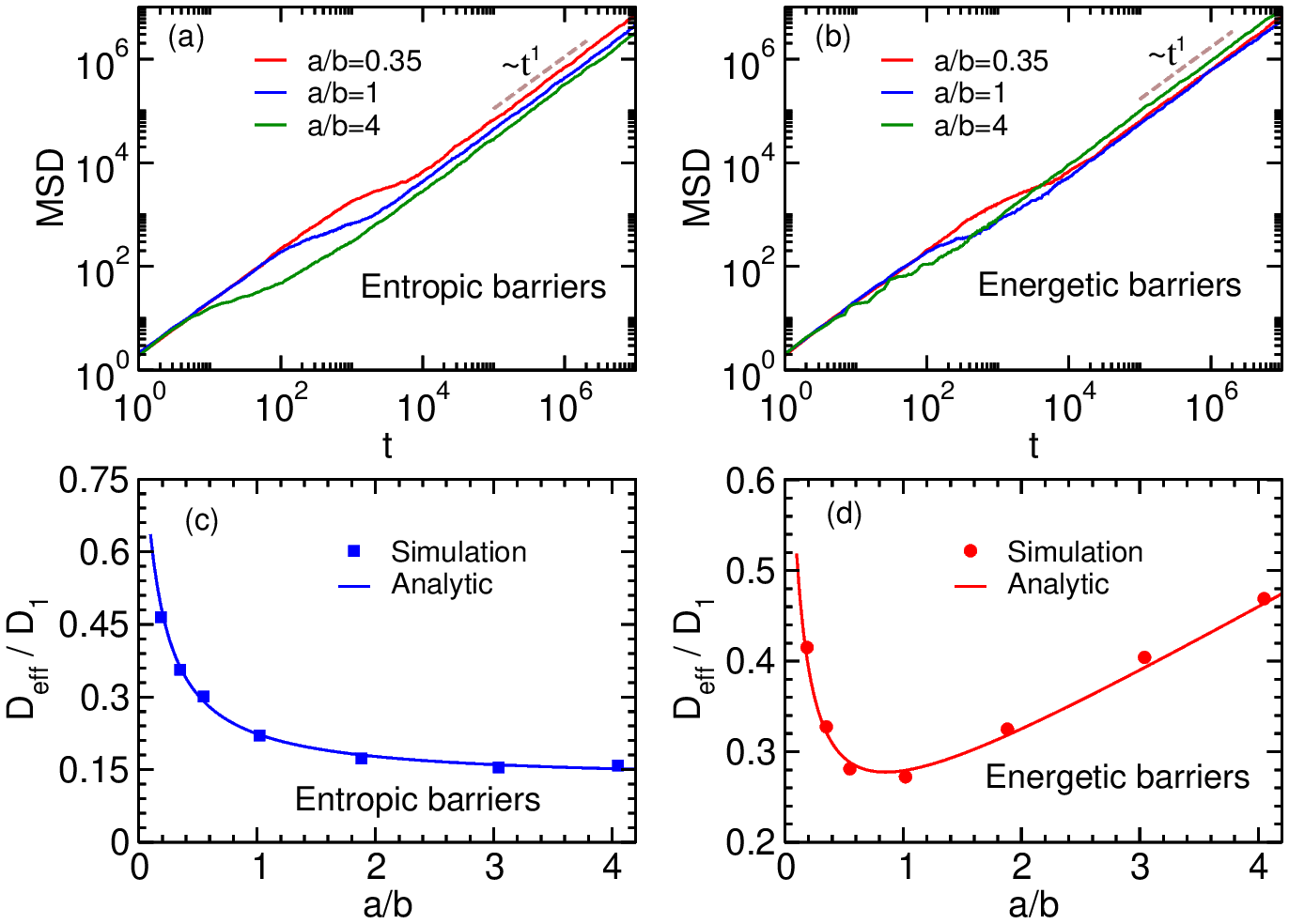}
    \caption{(a-b) Mean Squared Displacement (MSD) vs time is plotted with solid lines for diffusive particle in presence of entropic barriers and energetic barriers respectively (in 1D). The different colours denote different ratio of  the widths of barrier and empty region ($a/b$) and the dashed brown line is for the guidance of the eye. (c-d) Effective diffusion constant ($D_{\mathrm{eff}}$) is shown with variation of $a/b$ in case of entropic barriers and energetic barriers respectively. The points depict the simulation results while the solid lines denote the analytical result (Eq.~\ref{eq:Deff_entropic} for entropic barriers, and Eq.~\ref{eq:Deff_energetic} for energetic barriers).}
    \label{fig:Fig5}
\end{figure}

We now plot this MFPT from Eq.~\ref{eq:MFPT_impen_d} for $d=1,2$ with increasing $a/b$ (increasing barrier numbers) in Fig.~\ref{fig:Fig2}b. Remarkably, in all dimensions, the MFPT shows a non-monotonic behaviour, with a maximum time at a critical value of $a/b$. This implies that while initially energetic barriers slow down transport, beyond a critical number, increasing barriers decreases passage times, a counterintuitive finding driven by the nature of the {\em energetic barriers}. Note that unlike for entropic barriers, the scaled MFPT in this case depends on the dimensionality, with the scaled time being larger for $d=2$ compared to $d=1$. Again, similar to the discussion for entropic barrier, for $d>1$, there is an asymmetry depending on which boundary is absorbing (Fig \ref{fig:Fig2}b and Fig. S4 c,d). For the converse case, where the inner ($r=r_0$) boundary is absorbing, again the scaled time is lower than when $r=R$ is absorbing. However, the non-monotonic behaviour is a generic feature is all dimensions, implying that these {\em energetic barriers} regulate transport in a fundamentally different manner compared to entropic barriers.

\begin{figure*}[t]
    \centering
    \includegraphics[width=\linewidth]{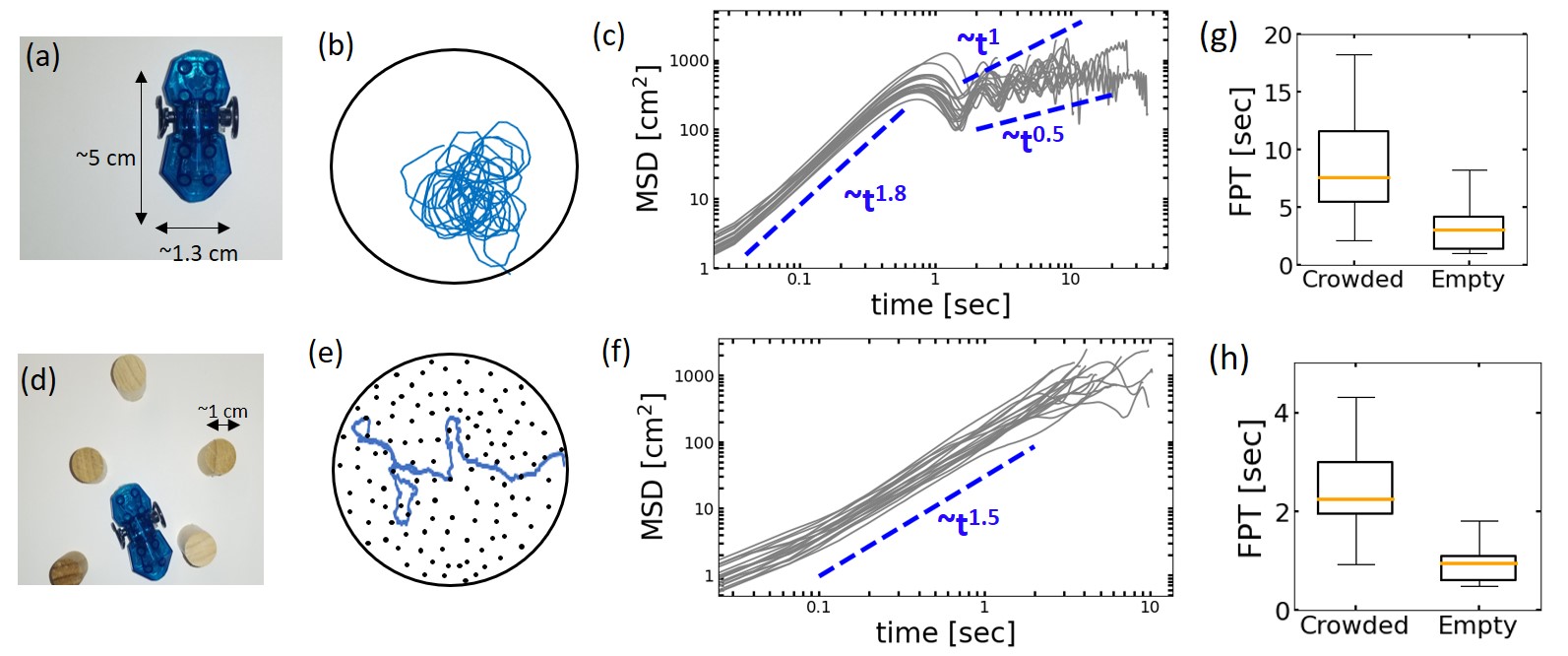}
    \caption{(a) A robotic bug (HEXBUG) is shown with scale bars. (b) Sample trajectory of a bug in empty circular space of diameter 117$cm$. (c) Mean square displacement of the bug as a function of time in the empty bulk space displaying a superdiffusive behavior (dotted line slope $\sim$ 1.8) for short times which demonstrates more diffusive and subdiffusive behavior at longer times with a fitted slopes between 0.5 and  1. (d) Robotic bug alongside wooden pegs with respective scalebar. (e) Sample trajectory of bug in crowded space of diameter 117$cm$ filled at a packing fraction of 6 $\%$. (f) Mean square displacement of bug through the crowded bulk area shows superdiffusive behavior with a fitted slope $\approx$ 1.5. (g) The characteristic mean first passage time of robotic bug in a crowded circular region (panel e) is higher than that in the empty region (h)The mean first passage time of robotic bug in a confined crowded rectangular region identical to barrier widths of the main experiments is recorded to be  higher than equivalent empty region.}
    \label{fig:Fig3}
\end{figure*}

Again in one-dimension (see SI Sec. 2.2.2 for details), Eq.~\ref{eq:MFPT_impen_d} simplies to give,\begin{equation}
    \frac{\tau}{\tau_0}=\frac{\left(1+\frac{a}{L}\right)^2+\left(\frac{a}{b}-\frac{a}{L}\right)\left(1+\frac{L}{a}\right)\frac{D_1}{Lv_q}}{(1+\frac{a}{b})^2}
\label{eq:scaled_mfpt_imp}
\end{equation}
Again this simple functional form makes evident the non-monotonic dependence of the MFPT on $a/b$. In $d=1$, the analytical result suggests that the maxima in MFPT occurs when the barrier width is comparable to the width of the empty regions, $a \sim b$. The location of the maxima can be calculated from Eq.~\ref{eq:scaled_mfpt_imp}, and yields, in the $L \rightarrow \infty$ limit, 
\begin{equation}
\left(\frac{a}{b}\right)^* = 1 - \frac{2a v_q}{D_1}
\end{equation}
By definition, $a/b\ge 0$, which implies $2a v_q/D_1 \le 1$. Moreover, if the hopping rate across a barrier is much lower than the bulk hopping rate, $q \ll p$, then $2a v_q/D_1 \ll 1$, implying that $(a/b)^* \simeq 1$.


\subsection*{Do signatures of non-trivial first passage persist in effective diffusivity?}
We now ask whether this conflicting behavior for the MFPT in the presence of entropic and energetic barriers has any signature on the more common transport properties of the system. Using kinetic simulations, we characterize the Mean Square Displacement (MSD) of the random walker as a function of elapsed time in an infinite lattice in the presence of barriers.  Note that, while the first passage property is history-dependent, the MSD is not.

For entropic barriers, the MSD is shown for three different $a/b$ ratios in Fig. ~\ref{fig:Fig5}(a). The RW initially explores the empty region in which it starts before it encounters the first barrier. This excursion is purely diffusive, with the bulk diffusion coefficient $D_1$, as is expected. At the timescale when it first encounters a barrier, the motion becomes subdiffusive as the barrier hinders the bulk diffusive behavior. Over long timescales ($t > 10^5$), the motion becomes diffusive again, however with an effective diffusion coefficient $D_{\mathrm{eff}}~i.e.~\avg{x^2(t)}=2 D_{\mathrm{eff}} t$. The value of $D_{\mathrm{eff}}$ is lower than the bulk value $D_1$. 

As $a/b$ increases, and $b$ decreases, with increasing $n$, the transition from early diffusive to a subdiffusive regime happens faster -- for $a/b = 0.35,1,4$, the crossover times are $t \sim 10^3, 10^2, 5$ respectively. Moreover, with increasing $a/b$ ratio the curves in Fig.~\ref{fig:Fig5}a at long times monotonically shift downwards. This in turn implies a monotonic decrease of $D_{\mathrm{eff}}$ as shown in Fig.~\ref{fig:Fig5}c. An analytical formula for $D_{\mathrm{eff}}$ exists in the literature \cite{garnett1904philtransroyalsoclondon,kalnin2012jchemphys} and can also be obtained by taking the $L \rightarrow \infty$ limit in Eq.~\ref{eq:scaled_MFPT} and expressing the MFPT as $\tau = L^2/2 D_{\mathrm{eff}}$, yielding 
\begin{equation}
    D_{\mathrm{eff}}= D_1 \left( \frac{a+b}{as+b} \right).
    \label{eq:Deff_entropic}
\end{equation}
The comparison on this analytical expression with the simulation is shown in Fig.~\ref{fig:Fig5}c. This monotonic behavior is consistent with the monotonic increase on the MFPT for entropic barriers in a finite domain. 

Next we turn to a similar characterization for the energetic barriers. The MSD of the RW on an infinite lattice, as above, is again shown for three different $a/b$ ratios in Fig.~\ref{fig:Fig5}b. Again, for all these case, there is an initial diffusive regime with a diffusivity $D_1$ of the empty regions. That crosses over to a subdiffusive regime when the RW starts to feel the effect of the barriers. As expected, this transition happens earlier  for the highest number of barriers ($a/b=4$), and later with decreasing $a/b$ ratios. In the long time limit, for all three $a/b$ ratios shown, the motions are again diffusive, with $\avg{x^2} = 2 D_{\mathrm{eff}} t$. However, quite strikingly in Fig.~\ref{fig:Fig5}d, the MSD of the intermediate barrier number (with $a/b=1$) lies below both the cases with lower and higher barrier numbers. As a result, as shown in Fig.~\ref{fig:Fig5}d, $D_{\mathrm{eff}}$ shows a non-monotonic behavior with increasing barrier number (or increasing $a/b$). Again, we can obtain an analytical expression for the $D_{\mathrm{eff}}$ by taking the $L \rightarrow \infty$ limit in Eq.~\ref{eq:scaled_mfpt_imp} and expressing the MFPT as $\tau = L^2/2 D_{\mathrm{eff}}$, yielding 
\begin{equation}
    D_{\mathrm{eff}}= D_1 \left( \frac{(1+\frac{a}{b})^2}{1+\frac{D_1}{b v_q}} \right).
    \label{eq:Deff_energetic}
\end{equation}
which matches the simulation results exactly, as shown in Fig.~\ref{fig:Fig5}d. Thus the signature of the non-monotonic dependence of the MFPT has its counterpart in the transport properties as well.



\section{Experimental Design}
We now design a simple experiment to probe this differential effect of energetic and entropic barriers in regulating first passage times. Given the generality of our results in any dimension, we design a simple quasi-1D experimental set up to verify our theoretical predictions. In our experiments, the random walker (RW) is a self-propelled robotic bug of dimension $5$ cm $\times$ $1.3$cm (see Fig.~\ref{fig:Fig3}a). These robotic bugs, called $hexbugs$, have been widely used in active matter research in experimental setups to demonstrate random motion that lead to interesting emergent individual and collective dynamics\cite{Boudet2021Science,Dauchot2019_PhysRevLett.122.068002,Dauchot2022_Nature}. In free space (see SI Sec. 1), we characterise the mean square displacement from multiple trajectories. A sample trajectory is shown in Fig.~\ref{fig:Fig3}b. The MSD reveals an initial superdiffusive behaviour, crossing over to a sub-diffusive behaviour at longer time scales, shown in Fig.~\ref{fig:Fig3}c. We use this RW to then study the behaviour in presence of entropic and energetic obstacles.

\begin{figure}[t!]
    \centering
    \includegraphics[width=\linewidth]{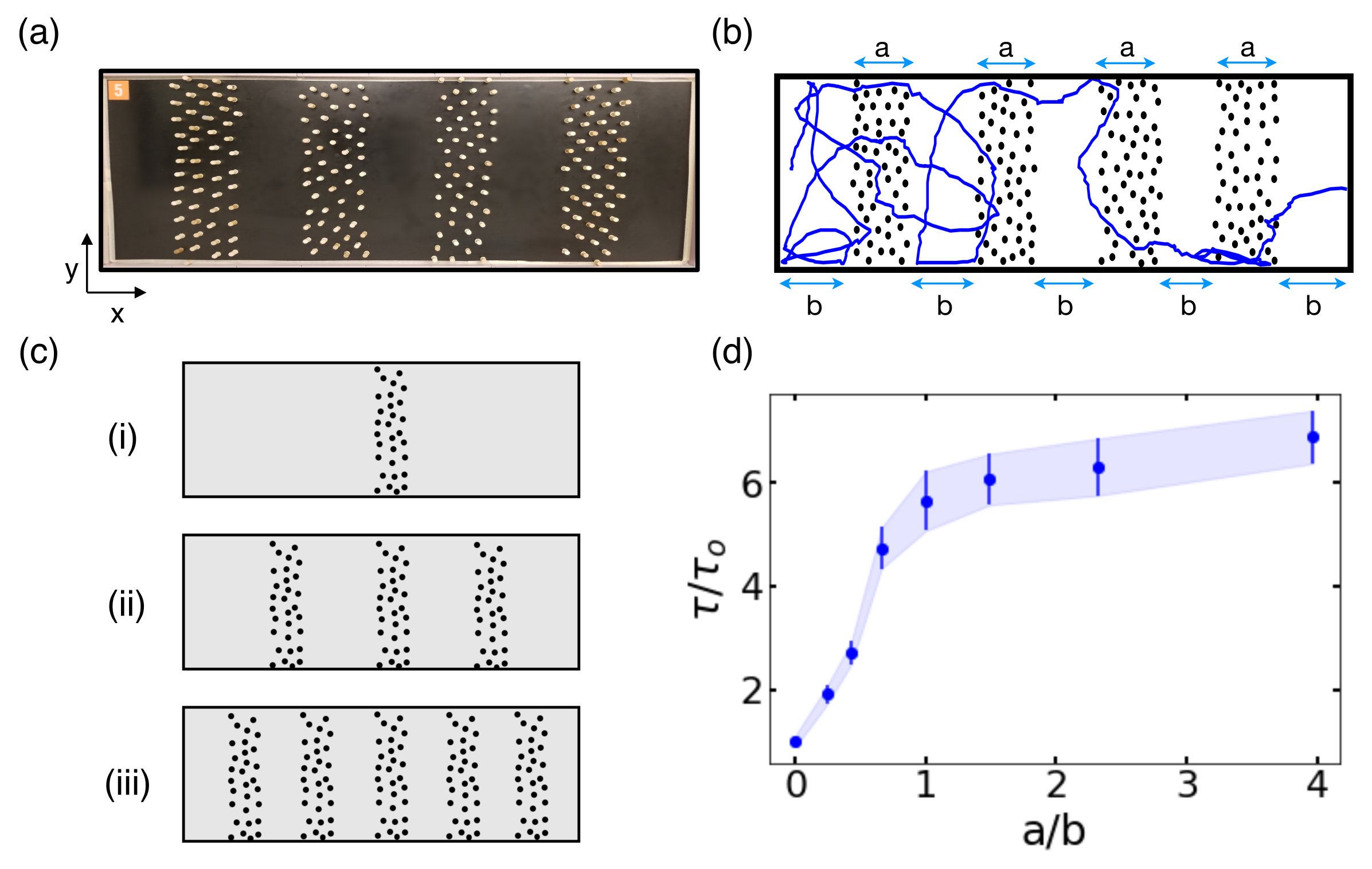}
    \caption{(a) Top view of the implemented 2D patterned geometry of alternate barrier and empty regions for entropic barriers. (b) Sample trajectory of bug this region. Barrier regions have a fixed width $a$, width of the empty region is denoted by variable $b$ (c) Schematic of arrangement of barriers for successive setups. We systematically increase the number of barriers to find the first passage time for each setup. (d) Scaled mean first passage time is plotted as a function of $a/b$. The MFPT increases monotonically with $a/b$, and saturates for high number of barriers. The error bars plotted in this curve are the standard errors ($SD/\sqrt{n}$)}.
    \label{fig:Fig4}
\end{figure}

We first focus on the case of entropic barriers. Our experimental setup consists of a rectangular region of dimensions $180 cm \times 76 cm$ having an array of $n$ obstacles alternating with $n+1$ free regions (Fig.~\ref{fig:Fig4}a). Each of these obstacles is a smaller rectangular strip of dimensions $20 cm \times 76 cm$, which are regions of lower diffusivity. To construct such regions, we used wooden pegs of diameter $1 cm$ arranged randomly at a packing fraction of $6\%$ to constitute a crowded barrier region Fig.~\ref{fig:Fig3}d. Our experimental premise of entropic barriers is different from other studies \cite{GhoshMetzlerNJP2016} which have characterized motion of diffusive walkers in two distinct ways : firstly the wooden pegs in our experiment are smaller compared to hexbug (walker) dimensions, secondly underlying motion of the hexbugs is
active, with a superdiffusive coefficient close to 1.5 (Fig.~\ref{fig:Fig3}f).  Bulk behavior of the hexbug in a crowded environment is depicted  in Fig.\ref{fig:Fig3}e,f.  The MFPT is about $2.5$ times larger than equivalent empty environment (Fig.~\ref{fig:Fig3}g), a trend which is also observed between crowded and empty rectangular barrier regions (Fig.~\ref{fig:Fig3}h). For any $n$, consistent with our theoretical framework, we are interested in the stochastic first passage of the bug starting from $x=0$ and reaching $x=L$ for the first time. The MFPT, $\tau (n)$, is then computed from $100$ experimental trials each for different barrier numbers $n$ (Fig.~\ref{fig:Fig4}c).

The behaviour of the MFPT is shown in Fig.~\ref{fig:Fig4}d as a function of the dimensionless ratio of the two widths of the barrier and free regions, namely $a/b$. The MFPT initially increases rapidly with increasing barrier numbers before saturating at high $a/b$ values. The behaviour of $\tau$ versus $a/b$ is monotonic as expected -- more the number of barriers, slower the navigation. Thus our experimental results validate the theoretical predictions in Eq.~\ref{eq:scaled_MFPT} for entropic barriers. 

\begin{figure}[t!]
    \centering
    \includegraphics[width=\linewidth]{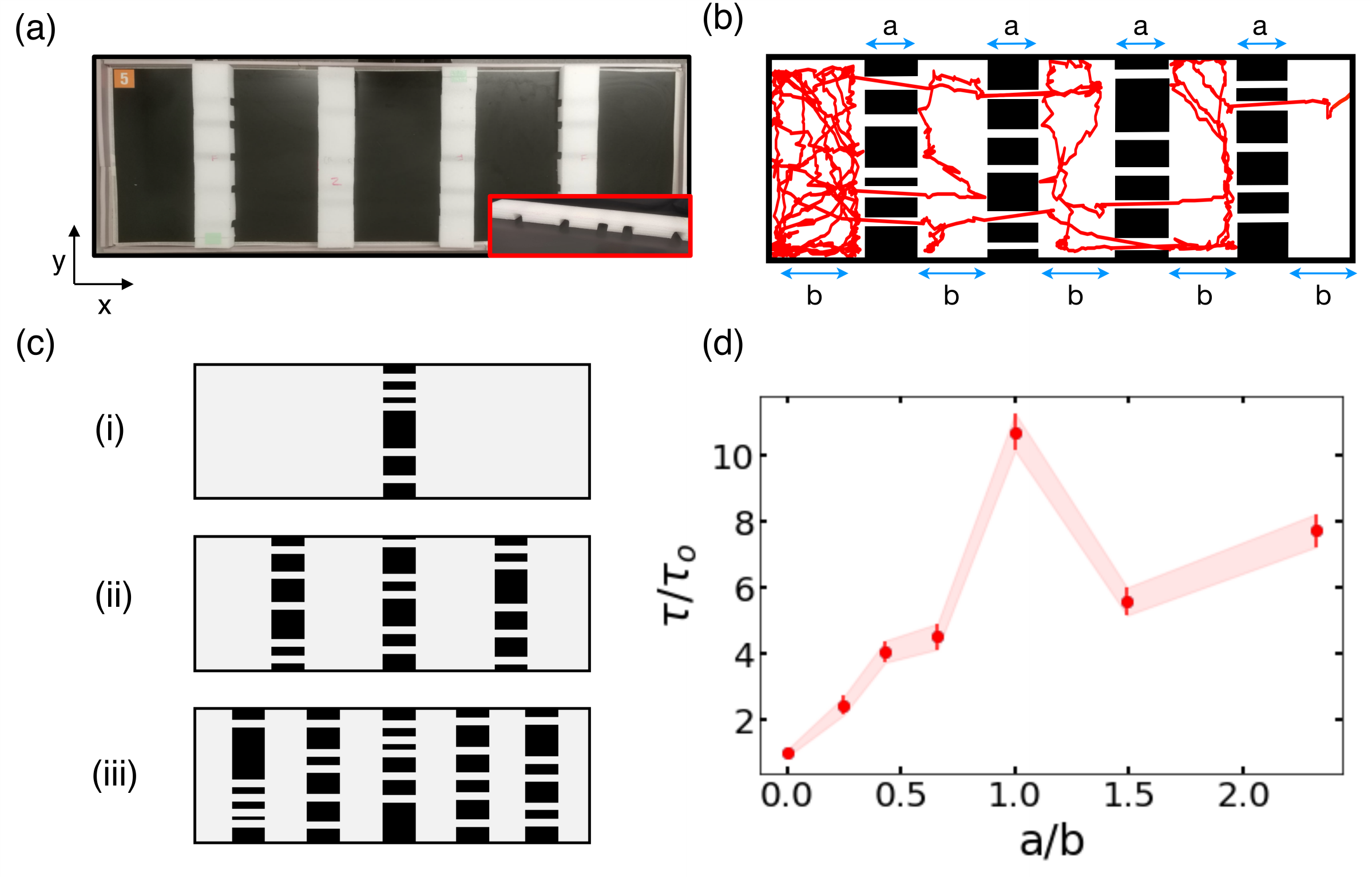}
    \caption{(a)Top view of implemented patterned geometry with alternate energetic barrier and empty regions. Inset depicts a side view of the styrofoam barrier. The individual tunnels are of size $3$cm. (b) Sample trajectory of bug through an arrangement of alternate empty (width $b$) and energetic barrier (of width $a$) regions. Trajectories show random motion of the bug in the empty regions, and an instantaneous motion through the tunnels whenever it finds them. (c)Schematic of arrangement of impenetrable barriers for successive setups, with different number of barriers. (d) Scaled MFPT is plotted as a function of $a/b$. The error bars plotted in this curve are the standard errors ($SD/\sqrt{n}$)}
    \label{fig:Fig6}
\end{figure}

We now construct obstacles to mimic the energetic barriers defined previously. We used styrofoam blocks of width $a = 20 cm$ as barriers with five tunnel-like passages of width $3 cm$ cut through them at random positions along their length (Fig.~\ref{fig:Fig6}a). A representative trajectory is shown in Fig.~\ref{fig:Fig6}b and successive barrier conformations are shown in Fig.~\ref{fig:Fig6}c. As the RW encounters the barrier, it most often gets reflected back into the free regions since the probability that it finds one of the five open tunnels at the right orientation is low. However, in contrast to the entropic barriers, once the bug enters into any of these tunnels, it passes through these regions very quickly ($\lesssim 1s$) (Fig. S2). Thus these barriers mimic the physics of energetic barriers, where the major delay comes from multiple barrier crossing attempts. 

We plot the behaviour of the MFPT $\tau$ for this array of energetic barriers as a function of the $a/b$ ratio in Fig.~\ref{fig:Fig6}d. For low number of barriers, the MFPT increases with increasing barrier number. However, as the $a/b$ ratio approaches the value $\sim 1$, the MFPT exhibits a local maximum. Increasing the number of barriers beyond this results in a decline in MFPT. This striking behaviour of a non-monotonic variation of MFPTs with increasing barrier numbers is in sharp contrast to the monotonic increase followed by saturation seen for entropic barriers, and is in line with our theoretical prediction in Eq. \ref{eq:scaled_mfpt_imp} for energetic barriers. The slight increase of the MFPT again at $a/b \simeq 2.5$ is an experimental design artefact (see SI - Sec. 1). Our experiments with two types of barriers thus suggest that the nature of the barriers play a crucial role in controlling the statistics of first passage times.

\color{black}

\section{Simulation of superdiffusive motion}
Our experiments suggest that this non-trivial effect of barriers in regulating first passage times is independent of the microscopic motion of the RW in the absence of barriers. The hexbugs in the experiments show directed motion at short time scales, and exhibit chiral motion at longer timescales in free space. The motion of these hexbugs broadly then falls with the regime of active matter \cite{ramaswamy2010annrevcondmattphys,romanczuk2012epjst,bechinger2016revmodphys}. Many physical, biological and even social system's dynamics can also be characterised by a similar non-diffusive transport. A few examples of active motion includes motion of flock of birds, school of fish, herds of land animals \cite{vicsek1995prl}, animal foraging \cite{viswanathan2011foraging}, E.coli bacteria \cite{javer2014natcomm}, janus particles \cite{zheng2013pre}, traffic \cite{helbing2001revmodphys} and many more.  Inside cells, superdiffusion has been observed for endogenous intracellular particles in crowded environments \cite{reverey2015scirep}. RNA polymerases and other motor proteins are also known to exhibit biased diffusion \cite{righini2018PNAS, rai2013cell}. The loop extrusion protein condensin is also known to exhibit superdiffusive motion on DNA backbones \cite{ganji2018science} and faces various protein obstacles which slows down loop extrusion \cite{pradhan2022cellrep}. A variety of theoretical frameworks have been used to model active particles, such as Run and Tumble particles (RTP) \cite{mori2020prl}, active Brownian particles (ABP) \cite{solon2015epjst}, and active Ornstein-Uhlenbeck particle (AOUP) \cite{martin2021pre}. Superdiffusive motion can also arise from non-markovian processes, as in the Continuous Time Random Walk (CTRW) \cite{metzler2014pccp} and fractional Brownian motion (fBm) \cite{mandelbrot1968siamrev}. Chiral active motion, as in the case of hexbugs, has also been observed experimentally in a variety of systems \cite{dauchot2019prl,deblais2018prl}, and investigated theoretically as well \cite{lowen2016epjst,liebchen2022epl}.

\begin{figure}[t!]
    \centering
    \includegraphics[width=\linewidth]{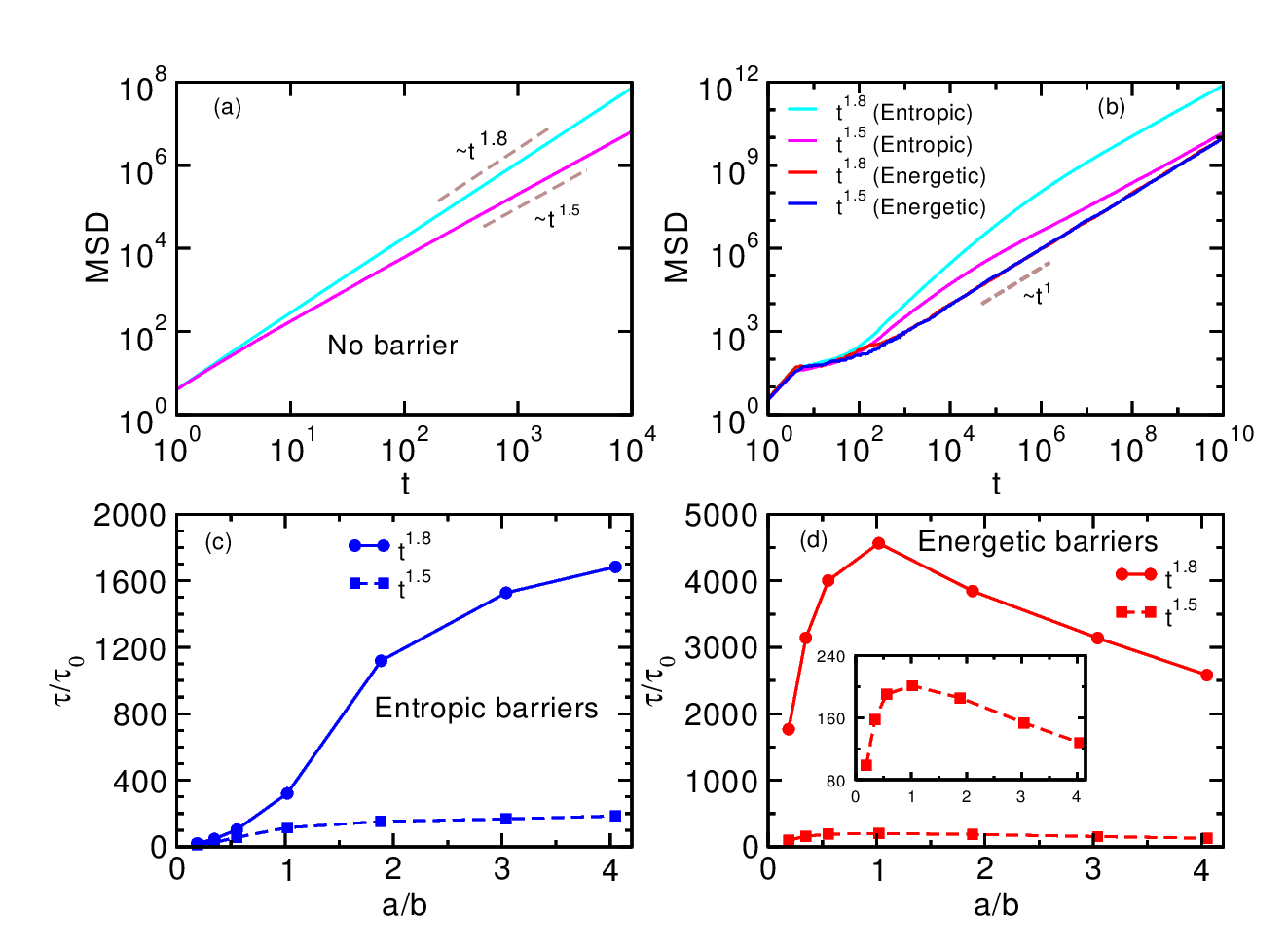}
    \caption{The figure depicts the motion of a superdiffusive particle in presence of barriers. Panel (a) shows the variation of MSD of the particle in absence of barriers while panel (b) shows the same in presence of barriers. Panels (c) and (d) denote the scaled MFPT vs $a/b$ for entropic and energetic barriers respectively. A zoomed in view of the dashed curve is shown in inset (panel (d)) to depict the non-monotonic behavior. }
    \label{fig:Fig7}
\end{figure}

In this section we investigate whether the non-trivial dependence of the barrier number on the first passage properties for entropic and energetic barriers is quite generic, and can be extended to systems with anomalous diffusion. To model superdiffusive transport for one-dimensional random walkers, we use a simple model known as the Elephant Random Walk (ERW) model \cite{schutz2004pre} that implements a non-markovian random walk with a full history-dependent memory with which it decides subsequent steps (see SI Sec. II for details). At any time, there is a probability $w$ with which it chooses its next step based on its history. It has been shown that the MSD in this case is superdiffusive: $\langle x^2_t \rangle \sim t^{4w-2}$ for $w > 3/4$ \cite{schutz2004pre}. We simulated the ERW for two values of $w = 0.875$ and $0.95$ for which the corresponding MSD exponents are $\alpha = 4w-2 = 1.5$ and $1.8$ respectively (see Fig. \ref{fig:Fig7}a). In the presence of barriers - both entropic and energetic - the MSD changes from an initial superdiffusive motion to a transient subdiffusive regime as it encounters the barriers (Fig. \ref{fig:Fig7}b). At longer times, the motion becomes diffusive, erasing the memory of the intrinsic superdiffusive motion in the barrier-free regions. This is shown for energetic barriers for the two superdiffusive walks in Fig. 6b. For entropic barriers, we observe the transition from an initial superdiffusive to the transient subdiffusive regime, and an approach to the limiting diffusive behavior.

On increasing the barrier numbers, the MFPT increases monotonically with the $a/b$ ratio for the case of entropic barriers, similar to the case of unbiased diffusion and our experiments. This is shown for both values of the superdiffusive exponent $\alpha$ in Fig. \ref{fig:Fig7}c. In contrast, for the case of energetic barriers, the MFPT again exhibits a non-monotonic behavior, similar to the case of ordinary diffusion and our experiments (Fig. \ref{fig:Fig7}d). The maximum of the MFPT occurs around $a/b \sim 1$, when the width of the barrier becomes equal to the length of the empty regions. Hence the differential impact of energetic and entropic barriers on first passage holds generically for different classes of RWs, and is thus very robust.


\section{Discussion}
We show that barriers may play an extremely non-trivial role in regulating first passage properties of systems. When barriers slow down transport, but allow lot of internal positional locations to be explored, we call them entropic barriers. In such cases, timescales of first passage across an array of barriers rise monotonically, as intuition would expect. Moreover the ratios of MFPT in the presence and absence of barriers saturate to a value greater than  one ($s=D_1/D_2 > 1$), which depends on the ratio of diffusion constants in the two regions.
On the other hand, when obstacles present a barrier that allows only rare entries (but near-instantaneous transport on successful entry) into the barrier regions, we call them energetic barriers.  In such cases, we show, using a combination of experiments and theory, that the timescales of first passage has a maximum when the length of the empty regions matches with the width of the barrier. This is also reflected in the behavior of the effective diffusivity, which has a corresponding minimum with increasing barrier density. This is in stark contrast to naive expectations that increasing barrier density should always result in slower times of passage.

We perform an exact analytical calculation for the mean first passage times in $d$ dimensions for both entropic and energetic barriers. For entropic barriers, at the barrier interfaces, the continuity of the MFPT and the currents (Eqs.~\ref{eq:left_T}-\ref{eq:right_dT}) allows us to derive a closed form expression for the MFPT which shows a monotonic increase with barrier numbers. For energetic barriers, the excluded barrier regions lead to a discontinuity in the mean times at the two barrier interfaces (Eq.~\ref{eq:T_2i-1_energetic}), which allows to derive a closed form solution which exhibits a non-monotonic behavior with barrier numbers. 

We also show that this curious behavior in the case of energetic barriers is not limited to pure diffusive transport alone, but also extends to situations where the transport is characterised by superdiffusion, within the framework of a simple non-Markovian model of transport. In general, this leaves open the question of how barriers affect the transport properties of different categories of active particles. Our results suggest that barriers may play an important role in regulating times of passage for superdiffusive transport, and this can be investigated through further experimental and theoretical studies. 

Regulation of first passage times are of critical importance in a variety of physical and biological systems. Our work highlights how differences in barrier design principles can crucially control times of passage in heterogeneous media.
We provide a generic experimental design principle which allows us to study these two cases - one in which there is an accessible but slow region of transport through the introduction of physical obstacles, and another in which we engineer tunnels which allow only rare barrier crossing events. It would be interesting to see whether these observations hold true for other possible implementations of barrier designs to mimic energetic and entropic barriers. Notwithstanding, these designs can serve as a template for future experimental studies to investigate the relative role of these two kinds of barriers in regulating effective transport properties.



\section*{Author Contributions}
M.D. designed and conducted the experimental study, analyzed data and wrote manuscript, S.G. carried out the analytical calculations and simulations, analyzed data and wrote manuscript, L.A. conducted the experimental study and analyzed data, D.D. analyzed data and wrote manuscript, M.K.M. designed the theoretical study, analyzed data and wrote manuscirpt.

\section*{Acknowledgements}
 We thank Kong Yang for help with preliminary data collection. M.D. acknowledges American Association of University Women (AAUW) Research Publication Grant in Engineering, Medicine and Science'21 for funding experimental aspects of this research. D.D. acknowledges SERB India (Grant No. MTR/2019/000341) for supporting this work. M.K.M. acknowledges support from SERB India (Grant No. CRG/2022/008142). S.G. acknowledges financial and computing support from IIT Bombay. 
\bibliography{main}
\bibliographystyle{unsrt}


\end{document}


\title{Nature of barriers determine first passage times in heterogeneous media}

\author{Moumita Dasgupta}
\email{dasgupta@augsburg.edu}
\thanks{These authors contributed equally}
\affiliation{Department of Physics, Augsburg University, MN 55454, USA}

\author{Sougata Guha}
\email{guha@na.infn.it}
\thanks{These authors contributed equally}
\affiliation{Department of Physics, IIT Bombay, Mumbai 400076, India}
\affiliation{INFN Napoli, Complesso Universitario di Monte S. Angelo, 80126 Napoli, Italy}

\author{Leon Armbruster}
\affiliation{Department of Physics, Augsburg University, MN 55454, USA}

\author{Dibyendu Das}
\email{dibyendu@phy.iitb.ac.in}
\affiliation{Department of Physics, IIT Bombay, Mumbai 400076, India}
\author{Mithun K. Mitra}
\email{mithun@phy.iitb.ac.in}
\affiliation{Department of Physics, IIT Bombay, Mumbai 400076, India}

\maketitle
\onecolumngrid

\vskip 100pt
\noindent 
\begin{center}
    {\Large{Supplementary Information}}
\end{center}
\newpage






\section{1. Experiment}

\begin{figure*}[h]
    \centering
    \includegraphics[width=0.7\linewidth]{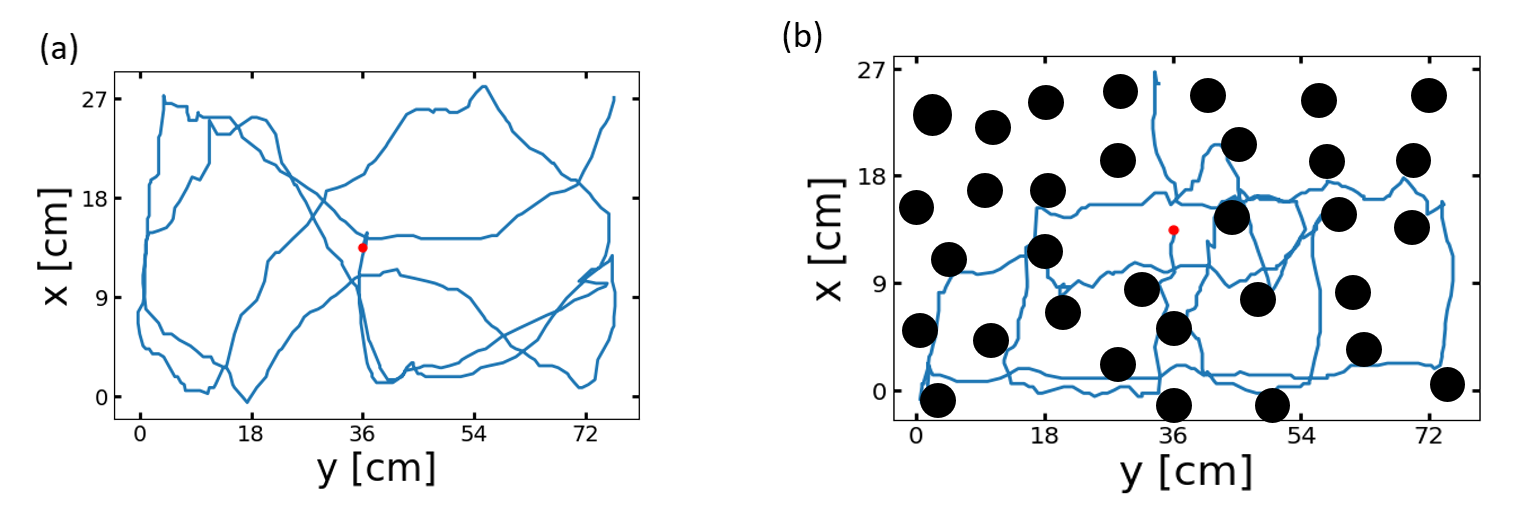}
    \caption{(a) Sample trajectory of RW in empty region of dimensions 76 $\times$ 27$cm$.  (b) Sample trajectory through obstacles of diameter 1 cm in a region of width 76 $\times$ 27 $cm$.}.
    \label{fig:S2}
\end{figure*}

A robotic bug (HEXBUG micro ant, $5cm \times 1.3 cm$) 
(Fig. 3a in main text) powered by battery is used as a self propelled random walker. We characterized the motion of our random walker (RW) in different setups. First we consider a relatively large circular confinement of diameter $117 cm$ (See Fig.$\sim$3b in main text). MSD (Fig.~1c in main text) are calculated for approximately 25 trajectories with the RW moving outward from the center of the confinement to the boundary. The MSD of the bug has a superdiffusive behavior for shorter timescale with t$\sim$1.8 which becomes subdiffusive over longer timescales with slopes between 0.5 and 1. To construct regions with a lower diffusion coefficient, we used wooden pegs of diameter $1 cm$ arranged randomly at a packing fraction of $6\%$ to constitute a crowded barrier region. A sample  trajectory and MSDs are shown in Fig.(3e,f) in main text. The MSD varies with a lower slope t$\sim$1.5 over a longer time compared to the empty case. The transport in this case is slowed down as compared to the free case due to repeated collision with the pegs, with the mean time to reach the boundary being $8.63 \pm 0.49 sec$, as compared to a mean time of $3.95 \pm 0.39 sec$ in the absence of barriers (Fig.3g in main text). The 6$\%$ packing fraction for the crowded region was determined experimentally based on the ease of turning of the robotic bug. Higher packing density resulted in the bug getting stuck between these crowders. We use this same density of pegs to construct rectangular barrier regions of dimensions $76 cm \times 27 cm$ (Fig.~\ref{fig:S2}). For these rectangular barriers, the MFPT is about $2.5$ times larger than for equivalent empty regions (Fig.3h in main text)

\begin{figure*}[h!]
    \centering
    \includegraphics[width=0.8\linewidth]{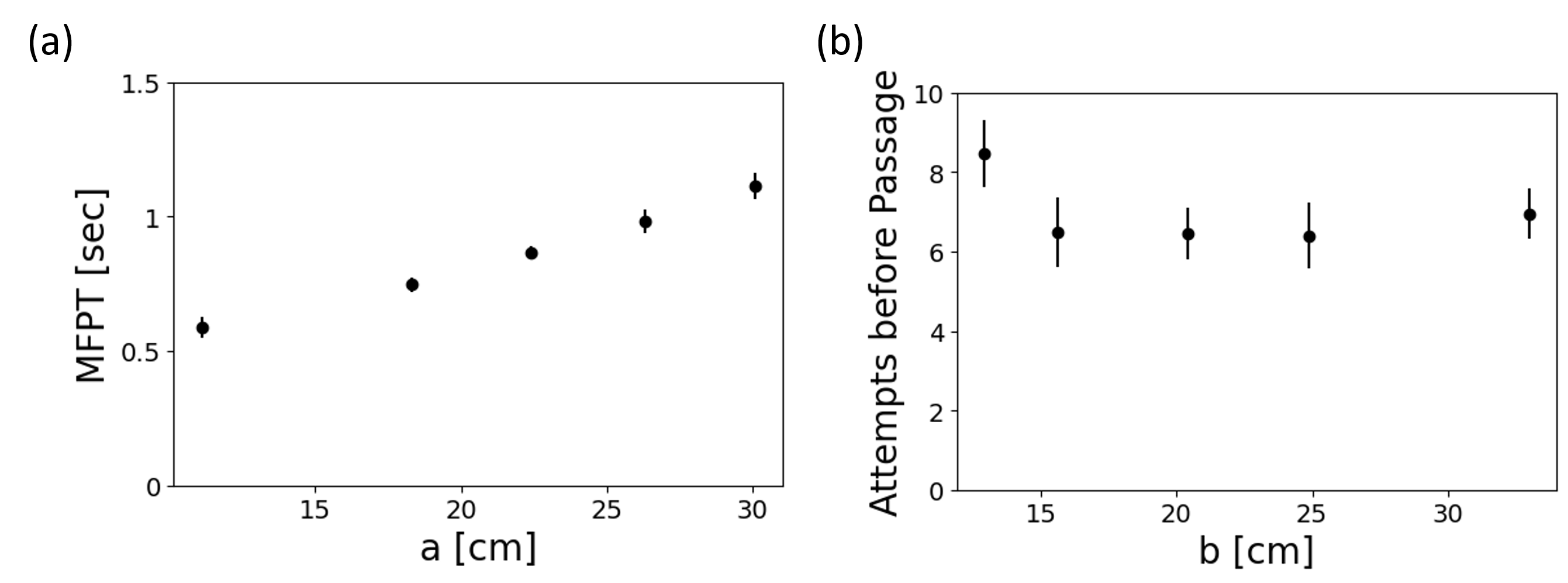}
    \caption{(a) Mean First Passage Time (MFPT) to traverse the barrier (tunnel) regions for the case of energetic barrier. The times increase roughly linearly with the width of the barrier region, indicating ballistic transport through the tunnels, as expected. (b) The number of times the RW collides with the barrier interface before it successfully enters the barrier (tunnel) region for the case of energetic barriers. This is a intrinsic property of the barrier, and is experimentally seems to be independent of the width of the barrier regions, as expected.}

    \label{fig:S3}
    \vspace{1in}
\end{figure*}

In our main experiments we had a rectangular length of dimension $180 cm \times 76 cm$ covered by an alternate barrier and empty regions. Two types of barrier regions are studied : entropic and energetic. For the entropic barriers, the pegs at 6$\%$ density filled a region of width $a$ = 20 cm. Energetic barriers were made  using styrofoams with 5 randomly-cut tunnels of size $3 cm$. This tunnel size (slight larger than the HEXBUG) allows the RW to pass through  in a near instantaneous manner (Suppl Fig.~\ref{fig:S3}(a)). For experimental barrier widths $a = 20 cm$, the passage times were $ \lesssim 1s$. Further, we also quantified the number of attempts before the RW successfully enters a tunnel, which is indicative of hopping probability accross the energetic barrier (Suppl Fig.~\ref{fig:S3}(b)).
Note that statistical fluctuations are quite high due to the fact that first passage time distributions in confined spaces are typically exponential tailed. As is expected for an exponential distribution, the average is comparable to the standard deviation. Therefore to obtain a reliable estimate of MFPT, an optimum number of trials had to be chosen. We observed that beyond 60-65 trials, the MFPTs were converging to a steady value. Hence we used 100 trials for each setup. 
For energetic barriers, the slight increase in MFPT beyond $a/b >2$ (Fig. 3d in main text), is an artefact of the finite size of our experimental setup. Around this width of the empty region ($b < 9$ cm), the dimensions of the RW become comparable to the width of the empty regions, and although the RW can still move, we do observe some significant restrictions in its turning behavior. The RW has a propensity to move preferentially only along the length of these narrow empty regions, without being able to turn randomly along the width - thereby increasing the overall MPFT. Since this behavior is not observed at any of the higher widths, this limitation of the system size is the cause for the higher MFPT values for $a/b > 2$.

\pagebreak

\section{2. Explicit analytic results}

\section{2.1 ABSORBING BOUNDARY AT $r=r_0$}

In the main manuscript we have derived MFPT for the case when absorbing boundary is located at $r=R$ and reflecting boundary at $r=r_0$. We now derive the opposite scenario where absorbing boundary is at $r=r_0$ and reflecting boundary is at $r=R$. Therefore, the boundary conditions in this case is given by,
\begin{equation}
    \left.\frac{\partial \avg{T_{2n+1}}}{\partial r}\right\vert_{R} = 0, \quad\quad \text{and} \quad\quad \left.\avg{T_{1}}\right\vert_{r_0}=0
\end{equation}

From the above Eq. it is easy to obtain,
\begin{eqnarray}
    A_{n+1} &=& -\frac{R^d}{d} \label{eq:A1}\\
    \text{and} \quad B_{1} &=& \begin{cases}
        -\frac{r_0^2}{2d} - \frac{A_{1}r_0^{2-d}}{2-d}, & \text{for $d\neq2$}\\
        -\frac{r_0^2}{2d} - A_{1}~ \ln r_0, & \text{for $d=2$}
    \end{cases} \label{eq:Bn1}
\end{eqnarray}

The matching conditions at barrier interfaces (Eqs. 6-9 and Eqs. 18-19 of the main text for entropic barriers and energetic barriers respectively) are independent of the choice of boundary conditions and hence the recursion relations given by Eqs. 12, 13, 14 and Eqs. 20, 21, 22 are still valid in this scenario.

Therefore, in this case , the MFPT of a diffusing particle starting from the reflecting boundary can be written as,
\begin{eqnarray}
    \tau = \avg{T_{2n+}1}\vert_{R} = \begin{cases}
        -\frac{1}{D_1}\left[\frac{R^2}{2d}+ \frac{A_{n+1}R^{2-d}}{2-d}+B_{n+1}\right], ~\text{for $d\neq2$} \\
         -\frac{1}{D_1}\left[\frac{R^2}{2d}+ A_{n+1}\ln R+B_{n+1}\right], ~ \text{for $d=2$} 
    \end{cases}
    \label{eq:MFPT_pen_d}
\end{eqnarray}
where for entropic barriers,
\begin{eqnarray}
 B_{n+1} = \begin{cases}
     B_{1} + \sum_{i=1}^{n}\frac{(s-1)}{2d}\left[r_{2i}^2-r_{2i-1}^2\right] \nonumber\\ + \sum_{i=1}^{n}\frac{(s-1)}{(2-d)}A_1 \left[r_{2i}^{2-d}-r_{2i-1}^{2-d}\right] \quad \text{for $d\neq2$} \nonumber\\
    B_{1} + \sum_{i=1}^{n}\frac{(s-1)}{2d}\left[r_{2i}^2-r_{2i-1}^2\right] \nonumber\\ + \sum_{i=1}^{n}(s-1)A_1 ~\ln\left(\frac{r_{2i}}{r_{2i-1}}\right) \quad\quad \text{for $d=2$} \nonumber
\end{cases}
\end{eqnarray}
and for energetic barriers,
\begin{eqnarray}
 B_{n+1} = \begin{cases}
     B_{1} - \sum_{i=1}^{n}\frac{1}{2d}\left[r_{2i}^2-r_{2i-1}^2\right] \nonumber\\ -\sum_{i=1}^{n}A_ir_{2i-1}^{1-d}\left[\frac{a_i}{(2-d)}-\frac{D_1}{v_q}\right] \nonumber\\+\sum_{i=1}^{n}\frac{a_ir_{2i}}{d(2-d)}+\sum_{i=1}^{n}\frac{D_1r_{2i-1}}{dv_q}, \quad \text{for $d\neq 2$} \\
    B_{n+1} -\sum_{i=1}^{n}\frac{1}{2d}\left[r_{2i}^2-r_{2i-1}^2\right]\nonumber\\ -\sum_{i=1}^{n}A_ir_{2i-1}^{1-d}\left[\frac{\ln r_{2i}}{r_{2i}^{1-d}}-\frac{\ln r_{2i-1}}{r_{2i-1}^{1-d}}-\frac{D_1}{v_q}\right] \nonumber\\ +\sum_{i=1}^{n}\frac{a_i\ln r_{2i}}{dr_{2i}^{1-d}} +\sum_{i=1}^{n}\frac{D_1r_{2i-1}}{dv_q}, ~ \text{for $d=2$} \nonumber
\end{cases}
\end{eqnarray}

\begin{figure}
    \centering
    \includegraphics[width=0.8\linewidth]{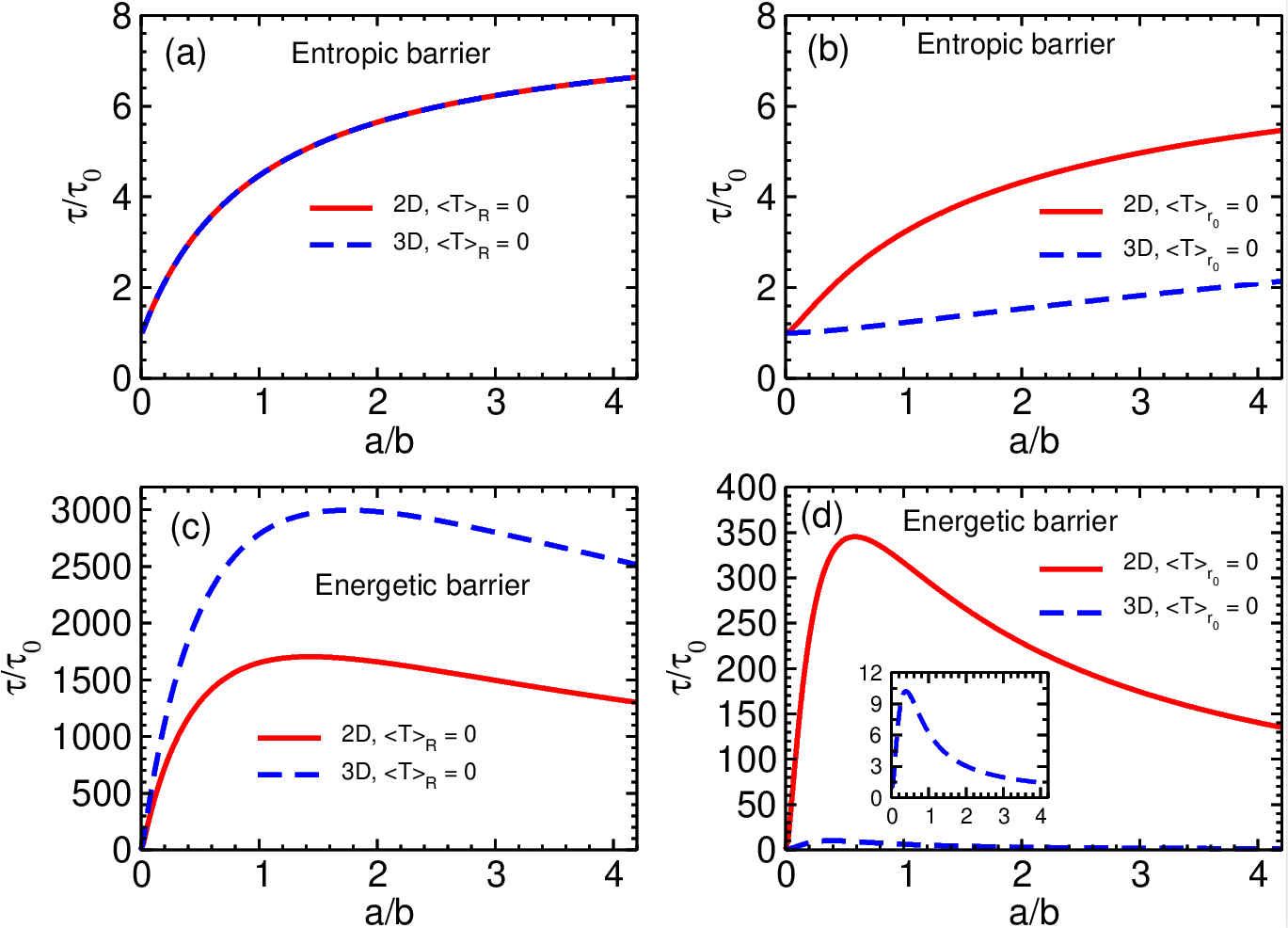}
    \caption{Comparison of MFPT in 2D and 3D for two opposite set up. Panels (a) and (b) compares the results for entropic barriers when the absorbing boundary is at $r=R$ and $r=r_0$ respectively for $D_1/D_2=8$. Panels (c) and (d) shows the results for $(D_1/dv_q)=50000$ when the absorbing boundary is at $r=R$ and $r=r_0$ respectively.  A zoomed in version of the MFPT curve in 3D for absorbing boundary at $r=r_0$ is shown in panel (d) inset. The other chosen parameters are $a=20$, $R=2001$ and $r_0=1$.}
    \label{fig:enter-label}
\end{figure}

\hskip 10in
\pagebreak

\section{2.2 Analytical calculations for one-dimension}
While we present the general analytical formalism in $d$-dimensions in the main manuscript, for simplicity, we also explicitly solve the 1D case here. 

\section{2.2.1  Entropic barriers}

Let us consider $n$ barriers each of width $a$ are distributed in region $x=0$ and $x=L$.\\
The mean gap between two consecutive barriers is given by
\begin{equation}
    b=\frac{L-na}{n+1}
    \label{eq:b}
\end{equation}

\begin{figure}[b!]
    \centering
    \includegraphics[width=\linewidth]{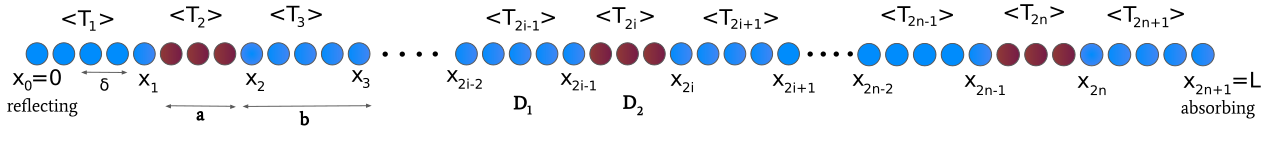}
    \caption{Schematic of the system with $n$ entropic barriers. The gap regions are denoted by blue and barriers are denoted by red. The boundary points of the $i^{th}$ barrier are denoted by $x_{2i-1}$ and $x_{2i}$. $\avg{T_i}$ denotes the mean average time for the particle to reach $x=L$ starting from the $i^{th}$ region.}
    \label{fig:analytical_schematic_entropic}
\end{figure}

The boundary points of $i^{th}$ barrier are denoted by $x_{2i-1}$ and $x_{2i}$ which can be expressed as
\begin{eqnarray}
x_{2i-1}&=&b+(i-1)(a+b) ~~~~~ i \in (1,n)\nonumber \\
x_{2i}&=&i(a+b)  ~~~~~~~~~~~~~~~~~~ i \in (1,n)
\label{eq:boundary_x}
\end{eqnarray}

The MFPT in two regions obey 
\begin{eqnarray}
D_1 \frac{\partial^2 \avg{T_{2i-1}}}{\partial x^2}&=&-1 ~~~~~~ i \in (1,n+1) \label{eq:diff1}\\
D_2 \frac{\partial^2 \avg{T_{2i}}}{\partial x^2}&=&-1 ~~~~~~ i \in (1,n) \label{eq:diff2}
\end{eqnarray}
which has solutions
\begin{eqnarray}
\avg{T_{2i-1}(x)}&=&-\frac{1}{D_1}\left[\frac{x^2}{2}+A_ix+B_i\right] ~~~~~~ i \in (1,n+1)\label{eq:diff_soln1}\\
\avg{T_{2i}(x)}&=&-\frac{1}{D_2}\left[\frac{x^2}{2}+C_ix+E_i\right] ~~~~~~ i \in (1,n)\label{eq:diff_soln2}
\end{eqnarray}
where $A_i,B_i,C_i,E_i$ are the integrating constants.\\
The boundary conditions of the lattice boundaries are given as,
\begin{eqnarray}
\partial_x<T_1>_{x=0}&=&0 ~~~~~~ \rightarrow \mathit{reflecting~boundary} \label{eq:ref_boundary}\\
<T_{2n+1}>_{x=L}&=&0 ~~~~~~ \rightarrow \mathit{absorbing~boundary} \label{eq:abs_boundary}
\end{eqnarray}
The continuity MFPT at $x_{2i-1}$ and $x_{2i}$ gives,
\begin{eqnarray}
\avg{T_{2i-1}}_{x_{2i-1}}&=&\avg{T_{2i}}_{x_{2i-1}} ~~~~~~~~~~ i \in (1,n) \label{eq:left_T}\\
\avg{T_{2i}}_{x_{2i}}&=&\avg{T_{2i+1}}_{x_{2i}} ~~~~~~~~~~ i \in (1,n) \label{eq:right_T}
\end{eqnarray}
Now the hopping dynamics that governs the motion can be written as,
\begin{eqnarray}
\avg{T_{2i-1}}_{x_{2i-1}} &=& \frac{1}{p+q}+\frac{p}{p+q}\avg{T_{2i-1}}_{-\delta+x_{2i-1}}+\frac{q}{p+q}\avg{T_{2i}}_{\delta+x_{2i-1}} \label{eq:hop_left}\\
\avg{T_{2i+1}}_{x_{2i}} &=& \frac{1}{p+q}+\frac{p}{p+q}\avg{T_{2i+1}}_{\delta+x_{2i}}+\frac{q}{p+q}\avg{T_{2i}}_{-\delta+x_{2i}} \label{eq:hop_right}
\end{eqnarray}
Now from Eq.\ref{eq:hop_left} we can write,
\begin{eqnarray}
(p+q)\avg{T_{2i-1}}_{x_{2i-1}} &=& 1+p\avg{T_{2i-1}}_{-\delta+x_{2i-1}}+q\avg{T_{2i}}_{\delta+x_{2i-1}} \nonumber\\
\Rightarrow p\left[\avg{T_{2i-1}}_{x_{2i-1}}-\avg{T_{2i-1}}_{-\delta+x_{2i-1}}\right] &=&1+q\left[\avg{T_{2i}}_{\delta+x_{2i-1}}-\avg{T_{2i-1}}_{x_{2i-1}}\right] \nonumber
\end{eqnarray}
Using Eq.\ref{eq:left_T} one can rewrite the above equation as,
\begin{eqnarray}
p\left[\avg{T_{2i-1}}_{x_{2i-1}}-\avg{T_{2i-1}}_{-\delta+x_{2i-1}}\right] &=&1+q\left[\avg{T_{2i}}_{\delta+x_{2i-1}}-\avg{T_{2i}}_{x_{2i-1}}\right] \nonumber\\
p\delta\left[\avg{T_{2i-1}}_{x_{2i-1}}-\avg{T_{2i-1}}_{-\delta+x_{2i-1}}\right] &=&\delta+q\delta\left[\avg{T_{2i}}_{\delta+x_{2i-1}}-\avg{T_{2i}}_{x_{2i-1}}\right] \nonumber
\end{eqnarray}
Now taking $\delta\rightarrow0$ we get,
\begin{eqnarray}
p\delta^2 \partial_x \avg{T_{2i-1}}_{x_{2i-1}} &=& q\delta^2 \partial_x \avg{T_{2i}}_{x_{2i-1}} \nonumber\\
\Rightarrow D_1\partial_x \avg{T_{2i-1}}_{x_{2i-1}} &=& D_2 \partial_x \avg{T_{2i}}_{x_{2i-1}} \label{eq:left_derivative}
\end{eqnarray}
Similarly from Eq.\ref{eq:right_T} and Eq.\ref{eq:hop_right} one can write,
\begin{eqnarray}
D_2\partial_x\avg{T_{2i}}_{x_{2i}}&=&D_1\partial_x<T_{2i+1}>_{x_{2i}} \label{eq:right_derivative}
\end{eqnarray}
Eqs.\ref{eq:ref_boundary},\ref{eq:abs_boundary},\ref{eq:left_derivative} and \ref{eq:right_derivative} constitutes the boundary conditions for Eqs.\ref{eq:diff_soln1} and \ref{eq:diff_soln2}. From Eq.\ref{eq:ref_boundary} we have,
\begin{equation}
    A_1 = 0 \label{eq:A1}
\end{equation}
Now, using Eq.\ref{eq:left_derivative} and Eq.\ref{eq:right_derivative} respectively we have,
\begin{eqnarray}
     C_i = A_i \label{eq:rCi_Ai}\\
     A_{i+1}=A_i \label{Ai}
\end{eqnarray}
Combining the Eqs.\ref{eq:A1},\ref{Ai},\ref{eq:rCi_Ai} we get,
\begin{equation}
    A_i=C_i=0 ~~~~~~\forall i=\{1,n+1\} \label{eq:recursion_A}
\end{equation}
Now from Eq.\ref{eq:left_T} we have,
\begin{eqnarray}
B_i&=&(s-1)\frac{x_{2i-1}^2}{2}+sE_i \label{eq:Bi_rEi}
\end{eqnarray}
where $s=D_1/D_2$.\\
Using Eq.\ref{eq:right_T} we have,
\begin{equation}
B_{i+1}=(s-1)\frac{x_{2i}^2}{2}+sE_i \label{Bi+1_rEi}
\end{equation}
Combining Eqs.\ref{eq:Bi_rEi} and \ref{Bi+1_rEi} we get,
\begin{equation}
    B_{i+1}=B_{i}+\frac{(s-1)}{2}\left[x_{2i}^2-x_{2i-1}^2\right] \label{eq:recursion_B}
\end{equation}
Now from Eq.\ref{eq:abs_boundary} we have,
\begin{equation}
    B_{n+1}=-\frac{x_{2n+1}^2}{2}=-\frac{L^2}{2} \label{eq:B_n+1}
\end{equation}
Hence using Eq.\ref{eq:recursion_B} we can write,
\begin{equation}
    B_1=B_{n+1}-\frac{(s-1)}{2}\sum_{i=1}^n \left(x_{2i}^2-x_{2i-1}^2\right) \label{eq:B_1}
\end{equation}
Therefore the MFPT of the particle to reach $x=L$ starting from $x=0$ in presence of $n$ barriers of width $a$ is given by,
\begin{eqnarray}
    \tau&=&\avg{T_1}_{x=0}=-\frac{B_1}{D_1} \nonumber\\
    &=& \frac{1}{D_1}\left[\frac{L^2}{2} + \frac{(s-1)}{2}\sum_{i=1}^n \left(x_{2i}^2-x_{2i-1}^2\right)\right] \label{eq:MFPT}
\end{eqnarray}
The summation in the above equation can be computed very easily.
\begin{eqnarray}
\sum_{i=1}^n \left(x_{2i}^2-x_{2i-1}^2\right)&=&\sum_{i=1}^{n}\left[i^2(a+b)^2-\{b+(i-1)(a+b)\}^2\right] \nonumber\\
&=&\sum_{i=1}^{n}\left[i^2(a+b)^2-\{i(a+b)-a\}^2\right] \nonumber\\
&=&\sum_{i=1}^n \left[2ia(a+b)-a^2\right] \nonumber\\
&=&n(n+1)a(a+b)-na^2 = naL \label{eq:summation}
\end{eqnarray}
If $\tau_0$ denotes the MFPT of the particle in absence of barriers then we have
\begin{equation}
    \tau_0=\frac{L^2}{2D_1}
\end{equation}
Therefore the scaled MFPT in presence of barriers is given by,
\begin{eqnarray}
\frac{\tau}{\tau_0}&=&\frac{2}{L^2}\left[\frac{L^2}{2}+ \frac{(s-1)}{2}\sum_{i=1}^n \left(x_{2i}^2-x_{2i-1}^2\right)\right] \nonumber\\
&=&1+\frac{(s-1)}{L^2}\cdot naL \nonumber\\
&=&1+(s-1)\cdot \frac{na}{L} \nonumber\\
&=&1+(s-1)\left(\frac{L-b}{a+b}\right)\frac{a}{L} \nonumber\\
&=&1+(s-1)\frac{L\left(1-\frac{b}{L}\right)}{b\left(1+\frac{a}{b}\right)} \cdot \frac{a}{L} \nonumber\\
&=&1+\frac{(s-1)\left(\frac{a}{b}-\frac{a}{L}\right)}{1+\frac{a}{b}}
\end{eqnarray}


\section{2.2.2  Energetic barriers}

In the same spirit we shall now solve first passage time in case of energetic barriers. The hopping dynamics that determines the motion can be written as,
\begin{eqnarray}
\avg{T_{2i-1}}_{x_{2i-1}}&=&\frac{1}{p+q}+\frac{p}{p+q}\avg{T_{2i-1}}_{-\delta+x_{2i-1}}+\frac{q}{p+q}\avg{T_{2i+1}}_{x_{2i}}\nonumber\\
\therefore p\left(\avg{T_{2i-1}}_{x_{2i-1}}-\avg{T_{2i-1}}_{-\delta+x_{2i-1}} \right)&=&1+q\avg{T_{2i+1}}_{x_{2i}}-q\avg{T_{2i-1}}_{x_{2i-1}}\nonumber\\
\avg{T_{2i-1}}_{x_{2i-1}}&=&\frac{1}{q}+\avg{T_{2i+1}}_{x_{2i}}-\frac{p\delta}{q}\partial_x\avg{T_{2i-1}}_{x_{2i-1}} \label{eq:T_imp_2i-1}
\end{eqnarray}
\begin{figure}[b!]
    \centering
    \includegraphics[width=\linewidth]{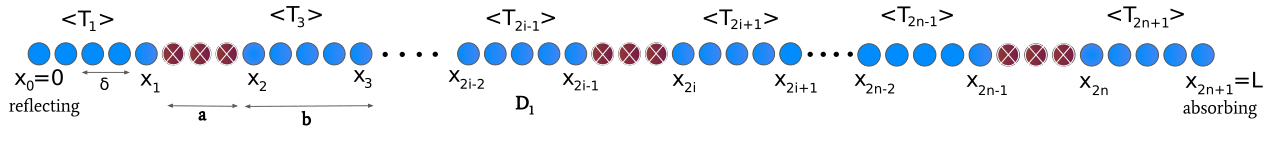}
    \caption{Schematic of the system with $n$ energetic barriers. The free regions are denoted by blue and barriers are denoted by red. The boundary points of the $i^{th}$ barrier are denoted by $x_{2i-1}$ and $x_{2i}$. $\avg{T_{2i-1}}$ denotes the mean  time for the particle to reach $x=L$ starting from the $i^{th}$ free region. Note that in this case $\avg{T_{2i}}$'s do not exist for all values of $i$.}
    \label{fig:analytical_schematic_eneregetic}
\end{figure}
Now we define $D_1=p\delta^2$ and $v_q=q\delta$ such that both $D_1$ and $v_q$ are finite at $\delta \rightarrow 0$. Therefore we must have $p,q\rightarrow \infty$ and thus Eq.\ref{eq:T_imp_2i-1} can be written as,
\begin{equation}
    \avg{T_{2i-1}}_{x_{2i-1}}=\avg{T_{2i+1}}_{x_{2i}}-\frac{D_1}{v_q}\partial_x\avg{T_{2i-1}}_{x_{2i-1}} \label{eq:T_x2i_1}
\end{equation}
Similarly it is easy to show that
\begin{equation}
    \avg{T_{2i-1}}_{x_{2i-1}}=\avg{T_{2i+1}}_{x_{2i}}-\frac{D_1}{v_q}\partial_x\avg{T_{2i+1}}_{x_{2i}} \label{eq:T_x_2i}
\end{equation}
Hence from Eq.\ref{eq:T_x2i_1} and Eq.\ref{eq:T_x_2i} we get,
\begin{equation}
    \partial_x \avg{T_{2i-1}}_{x_{2i-1}}=\partial_x\avg{T_{2i+1}}_{x_{2i}} \label{eq:equal_derivative}
\end{equation}
For n energetic barriers, there will be n+1 fast regions where the particle diffuses. Therefore,
\begin{equation}
    \avg{T_{2i-1}}=-\frac{1}{D_1}\left[\frac{x^2}{2}+A_ix+B_i \right] 
\end{equation}
where $i \in \{1,n+1\}$ denotes the region with boundaries $x_{2i-2}$ and $x_{2i-1}$ where the particle is in.\\
Now the lattice boundary conditions are same as Eq.\ref{eq:ref_boundary} and Eq.\ref{eq:abs_boundary},
\begin{eqnarray}
\partial_x \avg{T_1}_{x=0}=0 \label{eq:x0_condition}\\
\avg{T_{2n+1}}_{x=L}=0 \label{eq:xL_condition}
\end{eqnarray}
Now Eq.\ref{eq:x0_condition} gives
\begin{equation*}
    A_1=0
\end{equation*}
From Eq.\ref{eq:equal_derivative} we get,
\begin{equation}
    A_{i+1}=A_i-a = A_1-ia=-ia
\end{equation}
Using Eq.\ref{eq:xL_condition} we find $B_{n+1}$ as,
\begin{equation}
    B_{n+1}=-\frac{x_{2n+1}^2}{2}+nax_{2n+1} \label{eq:B_n+1_energetic}
\end{equation}
where $x_{2n+1}=L=(n+1)(a+b)-a$.\\
Now from Eq.\ref{eq:T_x2i_1} we can write,
\begin{equation}
    B_i=B_{i+1}+\frac{1}{2}\left(x_{2i}^2-x_{2i-1}^2\right)+\left(A_{i+1}x_{2i}-A_ix_{2i-1}\right)-\frac{D_1}{v_q}\left[x_{2i}+A_{i+1}\right]
\end{equation}
Therefore 
\begin{eqnarray}
B_1&=&B_{n+1}+\sum_{i=1}^n\frac{1}{2}\left(x_{2i}^2-x_{2i-1}^2\right)+\left(A_{i+1}x_{2i}-A_ix_{2i-1}\right)-\frac{D_1}{v_q}\left(x_{2i}+A_{i+1}\right) \nonumber\\
\Rightarrow B_1&=&B_{n+1}-\frac{n(n-1)}{2}a^2+\frac{n(n+1)D_1a}{2v_q}-\frac{n(n+1)}{2}\left(\frac{D_1}{v_q}+a\right)(a+b)+\frac{1}{2}\sum_{i=1}^n\left(x_{2i}^2-x_{2i-1}^2\right)\nonumber\\
\label{eq:B1}
\end{eqnarray}
Using Eq.\ref{eq:boundary_x}, Eq.\ref{eq:summation} and Eq.\ref{eq:B_n+1_energetic}, we can easily show from Eq.\ref{eq:B1} that
\begin{equation}
    B_1=-\frac{(n+1)^2b^2+n(n+1)\frac{D_1b}{v_q}}{2} \label{eq:B1_imp} 
\end{equation}
The scaled first passage passage time then can be written as,
\begin{eqnarray}
    \frac{\tau}{\tau_0}=-\frac{2B_1}{L^2} &=& \frac{(n+1)^2b^2+n(n+1)\frac{D_1b}{v_q}}{L^2} \nonumber
\end{eqnarray}
Now using the relation $L=(n+1)(a+b)-a=n(a+b)+b$ one can write,
\begin{eqnarray}
    \frac{\tau}{\tau_0}&=&\frac{\frac{(L+a)^2}{(a+b)^2}b^2+\left(\frac{L-b}{a+b}\right)\left(\frac{L+a}{a+b}\right)\frac{D_1b}{v_q}}{L^2} \nonumber\\
    &=&\frac{\frac{b^2(L+a)^2}{(a+b)^2}+\frac{(L^2+aL-bL-ab)}{(a+b)^2}\frac{D_1b}{v_q}}{L^2} \nonumber\\
    &=& \frac{\frac{b^2(L+a)^2}{L^2}+\frac{(L^2+aL-bL-ab)}{L^2}\frac{D_1b}{v_q}}{(a+b)^2} \nonumber\\
    &=&\frac{b^2\left(1+\frac{a}{L}\right)^2+\left(1+\frac{a}{L}-\frac{b}{L}-\frac{ab}{L^2}\right)\frac{D_1b}{v_q}}{(a+b)^2} \nonumber\\
    &=&\frac{b^2\left(1+\frac{a}{L}\right)^2+\left(\frac{L}{b}+\frac{a}{b}-1-\frac{a}{L}\right)\frac{D_1b^2}{Lv_q}}{(a+b)^2} \nonumber\\
    &=&\frac{b^2\left(1+\frac{a}{L}\right)^2+\left[\frac{a}{b}\left(\frac{L}{a}+1\right)-\frac{a}{L}\left(\frac{L}{a}+1\right)\right]\frac{D_1b^2}{Lv_q}}{b^2\left(1+\frac{a}{b}\right)^2} \nonumber\\
    &=&\frac{\left(1+\frac{a}{L}\right)^2+\left(\frac{a}{b}-\frac{a}{L}\right)\left(1+\frac{L}{a}\right)\frac{D}{Lv_q}}{\left(1+\frac{a}{b}\right)^2}
\end{eqnarray}

\section{3. Effective diffusivity}

In this section, we investigate the long time diffusivity in the presence of barriers. Using kinetic simulations, we characterize the Mean Square Displacement (MSD) of the random walker as a function of elapsed time in an infinite lattice in the presence of barriers.  Note that, while the first passage property is history-dependent, the MSD is not. 

For entropic barriers, the MSD is shown for three different $a/b$ ratios in Fig. \ref{fig:Fig5}a. The RW initially explores the empty region in which it starts before it encounters the first barrier. This excursion is purely diffusive, with the bulk diffusion coefficient $D_1$, as is expected. At the timescale when it first encounters a barrier, the motion becomes subdiffusive as the barrier hinders the bulk diffusive behavior. Over long timescales ($t > 10^5$), the motion becomes diffusive again, however with an effective diffusion coefficient $D_{\mathrm{eff}}~i.e.~\avg{x^2(t)}=2 D_{\mathrm{eff}} t$. The value of $D_{\mathrm{eff}}$ is lower than the bulk value $D_1$. 

As $a/b$ increases, and $b$ decreases, with increasing $n$, the transition from early diffusive to a subdiffusive regime happens faster -- for $a/b = 0.35,1,4$, the crossover times are $t \sim 10^3, 10^2, 5$ respectively. Moreover, with increasing $a/b$ ratio the curves in Fig.~\ref{fig:Fig5}a at long times monotonically shift downwards. This in turn implies a monotonic decrease of $D_{\mathrm{eff}}$ as shown in Fig.~\ref{fig:Fig5}c. In 1D, the MFPT of a free region (in absence of barriers) is given by $\tau_0=L^2/2D_1$. Analogously writing the MFPT of the heterogeneous medium as $\tau=L^2/2D_{\mathrm{eff}}$, it is easy to obtain the expression $D_{\mathrm{eff}}$ for the case $L\rightarrow \infty$. The Eq. 16 in the main text reads,
\begin{eqnarray}
    \frac{\tau}{\tau_0}&=& 1+\frac{(s-1)(\frac{a}{b}-\frac{a}{L})}{1+\frac{a}{b}} \nonumber\\
    \Rightarrow\quad \frac{L^2/2D_{\mathrm{eff}}}{L^2/2D_1}&=& 1+\frac{(s-1)\frac{a}{b}}{1+\frac{a}{b}} \quad\quad\quad [~a/L\rightarrow 0 \text{ as } L\rightarrow\infty~] \nonumber\\
    \Rightarrow\quad D_{\mathrm{eff}}&=&D_1\left(\frac{a+b}{as+b}\right)
\end{eqnarray}
The comparison of the simulation results with the effective diffusivity obtained from the theory is shown in Fig.~\ref{fig:Fig5}c. This monotonic behavior is consistent with the monotonic increase on the MFPT for entropic barriers in a finite domain.


Next we turn to a similar characterization for the energetic barriers. The MSD of the RW on an infinite lattice, as above, is again shown for three different $a/b$ ratios in Fig.~\ref{fig:Fig5}b. Again, for all these case, there is an initial diffusive regime with a diffusivity $D_1$ of the empty regions. That crosses over to a subdiffusive regime when the RW starts to feel the effect of the barriers. As expected, this transition happens earlier  for the highest number of barriers ($a/b=4$), and later with decreasing $a/b$ ratios. In the long time limit, for all three $a/b$ ratios shown, the motions are again diffusive, with $\avg{x^2} = 2 D_{\mathrm{eff}} t$. However, quite strikingly in Fig.~\ref{fig:Fig5}d, the MSD of the intermediate barrier number (with $a/b=1$) lies below both the cases with lower and higher barrier numbers. As a result, as shown in Fig.~\ref{fig:Fig5}d, $D_{\mathrm{eff}}$ shows a non-monotonic behavior with increasing barrier number (or increasing $a/b$). Again, following the same method as before, one can obtain $D_{\mathrm{eff}}$ from Eq. 23 of main text, 
\begin{eqnarray}
     \frac{\tau}{\tau_0}&=&\frac{\left(1+\frac{a}{L}\right)^2+\left(\frac{a}{b}-\frac{a}{L}\right)\left(\frac{1}{L}+\frac{1}{a}\right)\frac{D_1}{v_q}}{(1+\frac{a}{b})^2} \nonumber\\
     \Rightarrow \quad \frac{L^2/2D_1}{L^2/2D_{\mathrm{eff}}}&=&\frac{1+\frac{D_1}{bv_q}}{\left(1+\frac{a}{b}\right)^2} \quad\quad\quad [~1/L\rightarrow 0 \text{ as } L\rightarrow\infty~] \nonumber\\
     \Rightarrow \quad D_{\mathrm{eff}}&=&D_1\left(\frac{\left(1+\frac{a}{b}\right)^2}{1+\frac{D_1}{bv_q}}\right)
\end{eqnarray}
The comparison between the analytical expression with the simulation results is shown in Fig.~\ref{fig:Fig5}d. Thus the signature of the non-monotonic dependence of the MFPT has its counterpart in the transport properties as well.




\section{4. Superdiffusive motion}



To generate a driven motion of the particle we follow the \emph{Elephant-like memory diffusion} algorithm introduced by Sch\"utz and Trimper in 2004 \cite{schutz2004pre}. In this process the particle has complete memory of its previous steps. If at any time $t$ the particle is at $x_t$ then the evolution equation can be written as
\begin{equation*}
    x_{t+1}  = x_t + \sigma_{t+1}
\end{equation*}
where $\sigma_{t+1}$ is statistically chosen by the following method:
\begin{enumerate}
    \item First a previous timestep $t' \in \{1,2, ... , t\}$ is chosen randomly.
    \item Then, $\sigma_{t+1}=\sigma_{t'}$ with probability $w$ and $\sigma_{t+1}=-\sigma_{t'}$ with probability $1-w$.
\end{enumerate}
The particle starts at $x=0$ at $t=0$, and the first step of the particle is always towards positive direction in our simulation, i.e, $\sigma_1=+1$.

The MSD in this type of non-Markovian process follows \cite{schutz2004pre}:
\begin{eqnarray}
    \avg{x_t^2} &\sim& t \quad \quad \textrm{for} \quad w<0.75 \nonumber\\
    &\sim& t \; \textrm{ln} t \quad \textrm{for} \quad w=0.75 \nonumber\\
    &\sim& t^{4w-2} \quad \textrm{for} \quad w>0.75  \nonumber
\end{eqnarray}
For our simulations, we chose two values of $w = (0.875,0.95)$ in the regime $w > 0.75$ to recover superdiffusive transport, as mentioned in the text.

\color{black}
\bibliographystyle{unsrt} 
\bibliography{supplementary}